\documentclass[11pt]{article}
\usepackage{amsmath, amssymb}
\usepackage{amsthm}
\usepackage{graphicx}
\usepackage{geometry}
\usepackage{multirow}
\usepackage{bm}

\usepackage{latexsym, amssymb, amscd, amsxtra}
\usepackage{graphics, graphicx, color, subcaption, float, multirow}
\usepackage[dvipsnames]{xcolor}

\usepackage{fancyhdr}
\fancypagestyle{plain}{%
		\fancyhead[L]{\it to appear in Biometrika}
		
}

\usepackage[unicode=true,pdfusetitle, bookmarks=true,bookmarksnumbered=false,bookmarksopen=false,citecolor=blue, breaklinks=false,pdfborder={0 0 1},backref=false,colorlinks, urlcolor=blue]{hyperref}

\usepackage{natbib}

\providecommand{\lemmaname}{Lemma}
\providecommand{\propositionname}{Proposition}
\providecommand{\remarkname}{Remark}
\providecommand{\theoremname}{Theorem}

\usepackage{amsmath}               
  {
      \theoremstyle{plain}
      \newtheorem{assumption}{Assumption}
  }
  {
      \theoremstyle{plain}
      \newtheorem{proposition}{Proposition}
  }
  {
      \theoremstyle{plain}
      \newtheorem{remark}{Remark}
  }

\def\cH{\mathcal H}
\def\cI{\mathcal I}

\newcommand{\cov}{\mbox{cov}}
\newcommand{\E}{{E}}
\newcommand{\var}{\mbox{var}}
\renewcommand{\Pr}{\mbox{pr}}
\newcommand{\noprint}[1]{}

\title{Sparse Functional Linear Discriminant Analysis}

\author{Juhyun Park\thanks{Corresponding author: Juhyun Park,  
Department of Mathematics and Statistics, Lancaster University,
Lancaster LA1 4YF, U.K. and
ENSIIE \& LaMME, Universit\'e Paris-Saclay, 91025 \'Evry, France. 
email: juhyun.park@ensiie.fr}\\
Lancaster University, U.K. and ENSIIE, France\\
Jeongyoun Ahn \\
University of Georgia, U.S.A.\\
and Yongho Jeon\\
Yonsei University, South Korea}

\date{\today}

\begin{document}
\maketitle

\bibliographystyle{chicago}

\begin{abstract}
Functional linear discriminant analysis offers a simple yet efficient method for classification, with the possibility of achieving a perfect classification.
Several methods are proposed in the literature that mostly address the dimensionality of the problem. 
On the other hand, there is a growing interest in interpretability of the analysis, which favors a simple and sparse solution. 
In this work, we propose a new approach that incorporates a type of sparsity that identifies non-zero sub-domains in the functional setting, offering a solution that is easier to interpret without compromising performance. 
With the need to embed additional constraints in the solution, we reformulate the functional linear discriminant analysis as a regularization problem with an appropriate penalty. 
Inspired by the success of $\ell_1$-type regularization at inducing zero coefficients for scalar variables, we develop a new regularization method for functional linear discriminant analysis that incorporates an $L^1$-type penalty, $\int |f|$, to induce zero regions.
We demonstrate that our formulation has a well defined solution that contains zero regions, achieving a functional sparsity in the sense of domain selection. In addition, the misclassification probability of the regularized solution is shown to converge to the Bayes error if the data are Gaussian. Our method does not presume that the underlying function has zero regions in the domain, but produces a sparse estimator that consistently estimates the true function whether or not the latter is sparse.
Numerical comparisons with existing methods demonstrate this property in finite samples with both simulated and real data examples.

\end{abstract}

\section{Introduction} \label{sec:intro}

We consider a classification problem when data are curves. Denote the functional predictor by $X \in L^2(\cI)$, observable on a real interval $\cI$, and the class label by $Y$. Assuming that $X$ is a member of two possible groups ($Y=0$ or $Y=1$), 
we seek a classification rule that depends on a linear map, with unknown function $\beta$,
\begin{equation*}\label{e:Fbeta}
F_\beta(X) = \int_\cI X(t)\beta(t)\,\mbox{d}t .
\end{equation*}
This defines a functional linear classification problem, where we wish to determine $\beta(\cdot)$ in such a way that the linear map yields a good class separation. 
As functional data models are inherently infinite-dimensional, dimension reduction techniques are essential in constructing a solution. 

The standard logistic regression framework for classification can be extended with functional variables \citep{muller2005functional} as 
\[
\log\frac{\Pr(Y=1\mid X)}{\Pr(Y=0\mid X)} = a_0 + \int_\cI X(t)\beta(t)\,\mbox{d}t \,.
\]
Alternatively, one can try to directly extend the Bayes classifier. If $X$ were finite-dimensional, the Bayes rule is defined as maximizing the conditional probability given by 
\[
\Pr(Y=1\mid X) = \frac{\pi_1 f_1(X)}{\pi_0 f_0(X) + \pi_1 f_1(X)} \,,
\]  
where $i=0,1, \pi_i$ is the probability of $X$ coming from group $i$ and $f_i(X)$ is the marginal density of $X$ if it belongs to group $i$.
An optimal classification can then be achieved when one classifies new variable $X$ to the class 1 if $\pi_1 f_1(X) > \pi_0 f_0(X)$. Consequently, the main effort is devoted to estimating $f_1(X)$ and $f_0(X)$. If $X$ is Gaussian, further simplifications can be made and such an approach is broadly known as linear discriminant analysis. 
The main difficulty with extending the Bayes rule to functional data is linked to the fact that the marginal densities do not exist for functional data \citep{DelaigleHall2010}. Nevertheless, approaches attempting to directly approximate the marginal densities with dimension reduction techniques have proven useful, even in the absence of well-defined target densities. For example, \citet{JamesHastie2001} use a Gaussian framework to develop regularization methods, while \citet{BongiornoGoia2016} and \citet{DaiMuellerYao2017} develop nonparametric approaches without Gaussian assumptions. 

It is well known that functional classification can achieve a perfect classification, if the infinite-dimensionality is well exploited. 
This implies that for the purpose of classification, it is not necessarily advantageous to have a well-defined finite-dimensional representation. 
\citet{DelaigleHall2012} demonstrate such a phenomenon with a simple linear centroid classifier using asymptotic analysis and suggest a practical representation using components obtained from functional principal component analysis and partial least squares.  \citet{kraus2018classification} propose $L^2$ regularization methods to obtain the representation. \citet{BerrenderoCuevasTorrecilla2018} further clarify this phenomenon under a reproducing kernel Hilbert space framework for Gaussian processes, suggesting an alternative finite dimensional approximation.

While optimal performance is an important criterion to consider, the increasing impact of statistical analysis on modern scientific investigations has created the need to carefully consider interpretability of the outcomes of the analysis. 
Some attempts have been made to address interpretability in functional data, based on the idea that a simpler form of function is easier to interpret, and thus more useful in practice. 
As the function is an infinite-dimensional object, the formulation is often given in terms of a basis function representation.
Under this setting, three different approaches have been proposed to construct a simpler form of functions. The first one is to impose a constraint on the coefficients directly with an $\ell_1$ norm \citep[e.g.,][]{ZhouOgdenReiss2012}. Assuming that $\beta$ can be well approximated by a finite number of basis functions, say $K$, this can be expressed as $\beta(t) = \sum_{k=1}^K \alpha_k B_k(t)$ subject to $\sum_{k=1}^K |\alpha_j| \leq C$ 
and thus encourages the coefficients to be zero. 
The second approach is to limit the class of the functions $\{B_k\}$ in terms of their shape such as constant or strictly linear functions only \citep[e.g.,][]{TianJames2013}. 
A difficulty with a standard sparse regularization for a function is that the penalty that encourages a sparse representation of the function does not necessarily have a control over domain selection: i.e., even if $\alpha_k=0$ for some $k$, $\beta(t) \neq 0$ for $t \in I_k(t) = \{t \in \cI: B_k(t)\neq 0\}$.
The third one is to limit the support of the function $\beta$ to include zero regions \citep{JamesWangZhu2009, ZhouWangWang2013, MartinLilloRomo2014, LinWangCao2016, PichenyServienVilla2019}. 
We wish to incorporate interpretability in the latter notion of obtaining zero regions in the solution. 
However, as noted by \citet{KneipPossSarda2016} and \citet{Roche2018}, the theoretical framework appropriate to deal with a discrete notion of sparsity in high dimensions does not necessarily offer an insight into a problem in an infinite dimensional setting.

A functional formulation on sparsity is relatively scarce. \citet{WangKai2015} introduce the notion of functional sparsity, distinguishing global sparsity, which relates to functional variable selection, from local sparsity, which relates to domain selection with zero regions. \citet{TuParkWang2020} develop a regularization method to achieve simultaneous estimation of both types of sparsity in a time varying functional regression setting and \citet{LinCaoWangWang2017} propose an alternative regularization to achieve local sparsity in functional linear regression. Both approaches rely on a clever grouping of the sparse coefficients in the basis function representation. 
\citet{HallHooker2016} study the issue of identifiability of domain selection problem in functional linear regression and suggest a domain search strategy. \citet{kraus2018classification} follow a similar line. 

In this work, we seek an alternative approach to functional linear classification with a direct estimation method that  addresses dimensionality, optimality and interpretability.
With the need to embed additional constraints on the form of the solution, we reformulate the functional linear discriminant analysis as a regularization problem with an appropriate choice of penalty functions. 
So far, the penalty-based regularization methods have been used mostly for either smoothness in regression \citep[e.g.,][]{CardotFerratySarda2003, CrambesKneipSarda2009} or invertibility in classification \citep{kraus2018classification}, based on an $L^2$-type penalty.
The idea of sparsity as variable selection with an $\ell_1$-type penalty is actively developed in the high-dimensional setting with scalar variables, but much less for the infinite-dimensional functional setting.
We develop a new regularization method for functional linear discriminant analysis with an $L^1$-type penalty in the functional setting for inducing zero regions, as opposed to zero coefficients.  
We study the underlying optimization problem in detail, showing that it has a well defined solution that contains zero regions, achieving a functional sparsity in the sense of domain selection. 
Furthermore, we show that our regularized classifier asymptotically behaves similarly to the optimal classifier.
We do not presume, even in the case where the optimal classifier is well defined, that the underlying projection function $\beta$ has a local sparse property with true zero regions in the domain. The role of the $L^1$ penalty is to provide a sparse estimator that consistently estimates the true function whether or not the latter is sparse, enhancing interpretability without compromising prediction performance.
Unlike the approaches based on direct dimension reduction techniques such as principal component analysis or partial least squares, our proposed regularization method does not require assumptions on the eigenvalue sequences of the covariance function or any other unknown quantities.  
Numerical comparisons with existing methods demonstrate the properties of the proposed estimator in finite samples with both simulated and real data examples.

\section{Methodology} \label{sec:method}

\subsection{Functional data framework} \label{sec:fda}

We use the following standard notation: for $f\in L^p(\cI)$, $1\leq p<\infty$, the norm is defined by $\|f\|_p = (\int_\cI |f(t)|^p\,\mbox{d}t)^{1/p}$. 
When $p=\infty$, $\|f\|_\infty = \inf\{C: |f(t)|\leq C \mbox{ a.e.  on }\cI\}$.

Assume that $X \in L^2(\cI)$ with the mean $\mu(t)$ and covariance function $\cov\{X(s),X(t)\} = \gamma(s,t), s, t, \in \cI$. Define the inner product in $L^2(\cI)$ by $\langle f, g\rangle = \int_\cI fg$ for $f, g \in L^2(\cI)$. Assuming $\int_\cI\int_\cI \gamma(s,t)^2\,\mbox{d}s\mbox{d}t <\infty$, define the covariance operator $\Gamma: L^2(\cI) \rightarrow L^2(\cI)$ as
\[
\Gamma(\beta)(t) = \int_\cI \gamma(s,t)\beta(s)\,\mbox{d}s \,,\quad t \in \cI.
\]
It is known that $\Gamma$ is a compact operator, and is Hilbert-Schmidt, that admits a spectral representation given by
\[
\Gamma(\beta) = \sum_{j=1}^\infty \theta_j \langle \beta, \phi_j \rangle \phi_j \,,\qquad \theta_j \geq 0\,,\quad \theta_j \rightarrow 0 \,\,\mbox{ as } j \rightarrow\infty\,,\] 
where $\theta_j$ and $\phi_j(\cdot)$ correspond to eigenvalues  and eigenfunctions of the covariance operator $\Gamma$, respectively. One of the important properties of a compact operator is that it is not invertible unless it has only finitely many distinct eigenvalues.  
At the same time, since $\theta_j \rightarrow 0$ as $j \rightarrow\infty$, a finite rank approximation to $\Gamma$ is also well understood, which has been the basis of many regularization and dimension reduction techniques for functional data, notably functional principal component analysis and its applications \citep[e.g.,][]{HallPoskillPresnell2001}. More details on theoretical foundations of functional data are found in \citet{Hsing:Eubank:2015}.

For our classification problem, we denote the variable in each group by $X_0$ and $X_1$ and we make the following standard assumptions. 
\begin{assumption}\label{A:X} For $k = 0, 1$, $X_k$ is square integrable function on a compact interval $\cI$ with $\mu_k = \E(X_k)$ and $\mu_0\neq \mu_1$ with common covariance function $\gamma$. The mean functions and the covariance function are continuous.
\end{assumption}
\begin{assumption}\label{A:X4} For $k = 0, 1$, $\E(\|X_k\|_2^4) < \infty$.
\end{assumption}
Without loss of generality, we assume that $\mu_0 = 0$ and $\mu_1= \delta$. 

\subsection{An optimal linear classifier} \label{sec:optbeta}

The optimality of the linear methods in the homoskedastic Gaussian scenario is well established by \citet{DelaigleHall2012} based on an asymptotic centroid-based classifier. 
We briefly review their framework to motivate our regularization method, which is introduced in the following section.

For given $X$, the linear classifier with $\beta$ can be defined in multiple ways. We assume that $\beta \in L^2(\cI)$ is continuous.  The class assignment based on $\beta$ is done via
\[
T^0(X) = \big(\langle X, \beta\rangle  - \langle \delta, \beta\rangle\big)^2 - \big(\langle X, \beta \rangle\big)^2\,,
\]
which assigns to $X$ the group label $Y=1$ if $T^0(X) < 0$ and $Y=0$ if $T^0(X) >0$. 
Then, the misclassification error is $\pi_0 \Pr\{T_0(X)<0 \mid Y=0\} + \pi_1 \Pr\{T_0(X)> 0 \mid  Y=1\}$.
If $X$ is Gaussian, the misclassification probability can be expressed as 
\begin{equation}\label{e:error-gauss}
err_X(\beta) = 1 - \Phi\left(\frac{|\langle \delta, \beta\rangle|}{2\langle\beta, \Gamma \beta\rangle^{1/2}}\right) \,,
\end{equation}
with its minimal error given by $1 - \Phi(\|\Gamma^{-1/2}\delta\|_2/2)$ \citep{DelaigleHall2012, kraus2018classification}. 
It can be seen that an optimal function $\beta$ that minimizes the misclassification error is
\[
\max_{\beta\neq 0} \frac{\langle\delta,\beta\rangle^2}{\langle\Gamma\beta, \beta\rangle} =  \max_{\beta\neq 0}\frac{\langle \delta, \beta\rangle^2}{\var(\langle X, \beta\rangle)} \,.
\]
This is equivalent to the extended criterion for Fisher's discrimination analysis for functional data \citep{Shin2008} defined by 
\begin{equation}\label{e:fisher}
\max_{\beta\neq 0}\frac{\var[\E\{F_\beta(X)\mid Y\}]}{\E[\var\{F_\beta(X)\mid Y\}]} = \max_{\beta\neq 0} \frac{\pi_0\pi_1\langle\delta,\beta\rangle^2}{\langle\Gamma\beta, \beta\rangle}\,.
\end{equation}
Attempting to directly solve (\ref{e:fisher}) would lead to a form of generalized eigenvalue problem. 
Equivalently, this can be formulated as
\begin{equation}\label{e:lagrange}
\max_{\beta} \langle\delta,
\beta\rangle^2 \quad\mbox{ subject to }\quad  \langle\Gamma\beta,\, \beta \rangle - 1 = 0.
\end{equation}
The solution can be derived from the Lagrangian formulation of (\ref{e:lagrange})
\[
J_\rho(\beta) = \langle\delta,\,\beta\rangle^2 - \rho\{\langle\Gamma\beta,\,\beta\rangle - 1\} \,,
\]
for some constant $\rho$, which leads to
\begin{equation}\label{e:lda}
\Gamma\beta  = \delta  \,.
\end{equation} 
Hence, we see that the optimal linear classifier can be defined as the solution to (\ref{e:lda}). 
However, since a bounded inverse of $\Gamma$ does not exist in the infinite dimensional setting \citep[e.g.,][]{CardotMasSarda2007}, this equation clearly illustrates that linear discrimination analysis for functional data is an ill-posed inverse problem. When the covariance operator further satisfies 
$\|\Gamma^{-1}\delta\|_2 <\infty$,
a unique classifier exists in $(\mbox{Ker}(\Gamma))^\bot$ with the optimal error $1-\Phi(\|\Gamma^{-1/2}\delta\|_2/2)$. When $\|\Gamma^{-1}\delta\|_2 = \infty$, an optimal solution does not exist, but optimal classification can be achieved asymptotically along a non-convergent path and perfect classification may be possible if $\|\Gamma^{-1/2}\delta\|_2 = \infty$.
Therefore, unlike regression, it is not necessary to assume the existence of a unique solution to obtain an approximate solution, whose performance can mimic the optimal classifier asymptotically \citep{DelaigleHall2012, kraus2018classification}.  
In the following, we will denote the solution to (\ref{e:lda}) by $\beta_0$ when it exists and is finite. 

\subsection{A regularized solution to discrimination} \label{sec:penEst}

The primary purpose of regularization is to solve an ill-posed inverse problem. We introduce our regularization method to solve the ill-posed inverse problem (\ref{e:lda}) above. A sensible strategy would be to impose some constraints on the solution set or a penalty term to an objective function. In order to impose the functional equation (\ref{e:lda}), we introduce a corresponding minimization problem.   
Specifically, viewing the discriminant equation in (\ref{e:lda}) as a type of score equation leads to defining an objective function as
\begin{equation}\label{e:lda2}
\ell(\beta) 
=\frac{1}{2}\langle\Gamma\beta,\beta\rangle - \langle\delta,\beta\rangle \,.
\end{equation}
Hence the functional derivative of ~(\ref{e:lda2}) with respect to $\beta$ yields (\ref{e:lda}).
In practice, where $\Gamma$ and $\delta$ are not available, replacing $(\Gamma,\delta)$ by their empirical counterpart $(\Gamma_n,\delta_n)$ gives 
\[
\ell_n(\beta) = \frac{1}{2}\langle\Gamma_n\beta, \beta\rangle - \langle\delta_n, \beta\rangle \,.
\]
For example, the standard sample covariance operator and the sample mean could be used for $\Gamma_n$ and $\delta_n$. 
The standard sample estimators often lack smoothness \citep[e.g.,][]{CardotFerratySarda1999} so it is often desirable to use their smoothed version by taking into account the underlying design schemes \citep{ZhangWang2016}.  Our formulation does not rely on the specific choice of estimators as long as they are consistent.  
To incorporate additional constraints, we consider the following optimization problem
\begin{equation}\label{e:minPen}
\min_{\beta\in \cH} \{\ell_n(\beta) + P(\beta)\} \,,
\end{equation}
where $\cH$ is an appropriate function space and $P(\beta)$ is a penalty corresponding to the constraint. 
The type of constraint requires some further consideration. As the underlying problem is defined in $L^2(\cI)$, it seems natural to add a penalty that depends on the $L^2$ norm of $\beta$ such as $\|\beta\|_2^2$ as in \citet{kraus2018classification}, 
while a smoothness constraint on $\beta$ would lead to the $L^2$ penalty on the derivatives.   
Nevertheless, $L^2$ regularization in general does not produce a sparse solution \citep[e.g.,][]{LinCaoWangWang2017, TuParkWang2020}. This phenomenon is well investigated in multivariate data, and the popularity of the $\ell_1$ or $\ell_0$ penalty demonstrates the effectiveness of a sparse solution in a broader context.

In order to introduce a type of sparsity in the discriminant function, 
we propose to directly impose the functional norm constraints $\|\beta^\prime\|_2^2 \leq C_1$ and $\|\beta\|_1 \leq C_2$.
Although other constraints are possible, 
it turns out that this combination of constraints allows direct control over the $L^2$ norm. 
One might wonder whether the derivative penalty is necessary in our sparse regularization. Our experiences suggest that a combined norm give more stable results. 
We have assumed compact support to simplify the statement in Assumption~\ref{A:X}.
Under the assumption of compact support for $\beta$, the role of the derivative penalty does not seem so significant with respect to the classification performance. 
In general, since we only have $\|\beta\|_1 \leq \|\beta\|_2$, the bound on the $L^1$ norm alone is not sufficient to ensure the $L^2$ properties of $\beta$ and the techniques we have developed below do not necessarily work.
On the other hand, we have \citep[e.g.,][]{Gabushin1967, LiLeoni2018} that
\begin{equation*}
\|\beta\|_2 \leq \|\beta\|_1 + \|\beta^\prime\|_2 \,,
\end{equation*}
which suggests that the proposed method is able to control the $L^2$ norm of the function more effectively by regularizing both the derivative and the $L^1$ norm of the function.

Taking into account the functional constraints in (\ref{e:minPen}) leads to the following objective function
\begin{equation}\label{e:Jn}
J_n(\beta)=\frac{1}{2}\langle \Gamma_n\beta,\beta\rangle - \langle \delta_n,\beta\rangle + P(\beta) \,,
\end{equation}
where 
$P(\beta) = \lambda\|\beta\|_1 + (\eta/2)\|\beta^\prime\|_2^2$ and $\lambda$ and $\eta$ are tuning parameters, chosen from data. 

In this formulation, the derivative penalty is a standard smoothness penalty. The non-standard part is the $L^1$ penalty, which is an infinite-dimensional counterpart of the $\ell_1$ sparsity penalty. As is demonstrated below, the $L^1$ norm can be linked to a sparse solution in function space in terms of domain selection \citep{LiLeoni2018}, which is a special case of functional sparsity~\citep{WangKai2015}. Arguably, the result would be easier to interpret as the localized effects are automatically identified. 

We note that it is possible to use the ridge penalty ($\|\beta\|_2^2$) instead of the derivative norm. In fact, this choice makes the analysis of the optimization problem much easier as it directly controls the $L^2$ norm. In this case, the fact that $\langle\Gamma_n\beta, \beta\rangle + \tau \|\beta\|_2^2  \geq \tau \|\beta\|_2^2$ for all $\tau>0$ is sufficient to guarantee the existence of a solution. Nevertheless, we have chosen the derivative norm that has the added consideration of smoothness of the solution, since a smoother solution is easier to interpret.

\subsection{Properties of the regularized classifier in the population model} \label{sec:J1}

In this section we study the characteristics of the proposed regularized solution to the infinite-dimensional convex optimization problem. 
In order to simplify the discussion, we first consider the following optimization problem that replaces the sample quantities by the population counterparts.
Let  
\begin{equation}\label{e:J1}
J(\beta) = \frac{1}{2}\langle \Gamma\beta, \beta\rangle - \langle \delta,\beta\rangle + \lambda \|\beta\|_1 + \frac{\eta}{2}\|\beta^\prime\|_2^2 \,.
\end{equation}
The smoothness constraint with $\beta^\prime$ reduces the feasible set of $\beta$ from $L^2(\cI)$ to a differentiable subspace $H^1(\cI) = \{\beta\in L^2(\cI): \beta \mbox{ is absolutely continuous}, \beta^\prime \in L^2(\cI)\}$, which is the Sobolev space $H^1 = W^{1,2}$ in standard notation \citep{Brezis2011}. 
Since $L^2(\cI) \subset L^1(\cI)$, we consider $\cH = \{ \beta\in H^1(\cI): \|\beta\|_1 + \|\beta^\prime\|_2 < \infty\}$ as the space of feasible solutions, equipped with the norm given by $\|\beta\| = \|\beta\|_1 + \|\beta^\prime\|_2$. 
Then it is clear that $\cH$ is a convex set. 

Under this setting, it can be shown that $J$ is convex, so that the existence of a solution is guaranteed. Moreover, we can show that $J$ is strictly convex so the solution is unique. 
We summarize the result in the following proposition. 

\begin{proposition}\label{prop:existence} The function $J:\cH \rightarrow \mathbb{R}$ defined in (\ref{e:J1}) has a global minimizer $\tilde\beta \in \cH$ such that
\[
J(\tilde\beta) \leq J(\beta) 
\]
for all $\beta \in \cH$. Furthermore, the solution is unique. 
\end{proposition}

In order to understand the property of the solution $\tilde\beta$, additional characterizations of the solution beyond its existence are necessary. Note that if the objective function is differentiable, the first derivative (along with the second derivative) condition sufficiently characterizes the solution and a Newton-type algorithm can be used to find it. Non-differentiable functions require an alternative form to replace the derivative condition. Following similar arguments in \citet[ch 6]{Reyes2015} and \citet[p. 70-71]{Glowinski1984}, we derive a set of conditions that $\tilde\beta$ satisfies in the following proposition, which could be used for developing an optimization algorithm.

\begin{proposition}\label{prop:solution}  If $\tilde\beta$ is the minimizer of $J(\beta)$ in (\ref{e:J1}), there exists a function $\alpha \in L^2(\cI)$ with $|\alpha(t)|\leq \lambda$ a.e. for $t \in \cI$ such that $\tilde\beta$ satisfies the following relation:
\begin{equation}\label{e:sol1}
\langle\Gamma \tilde\beta, \beta\rangle + \eta \langle \tilde{\beta}^\prime, \beta^\prime\rangle - \langle\delta, \beta\rangle + \langle \alpha, \beta\rangle = 0 \,,\quad \mbox{ for all } \quad \beta\in \cH \,,
\end{equation}
and
\begin{equation}\label{e:sol-alpha}
\left\{\begin{array}{ll}  \alpha(t) = \lambda & \mbox{ on } \{t \in \cI: \tilde\beta(t)>0\}\,,  \\
|\alpha(t)| \leq \lambda  & \mbox{ on } \{t \in \cI: \tilde\beta(t)=0\}\,, \\
\alpha(t) = -\lambda & \mbox{ on } \{t \in \cI: \tilde\beta(t)<0\} \,. \end{array}\right.
\end{equation}
\end{proposition}

\medskip

The equations (\ref{e:sol1}) and (\ref{e:sol-alpha}) are necessary conditions for the solution to satisfy. 
Equation (\ref{e:sol1}) means that the subgradient of the objective at the minimizer $\tilde\beta$ contains zero. 
Furthermore, Proposition~\ref{prop:solution} demonstrates the domain selection property of the solution.
In order to make an analogy to the finite high dimensional case, consider a simplistic example where $\Gamma = I$, the identity operator, and $\eta=0$ in (\ref{e:J1}). Note that, since the identity operator is not compact, this is not a realistic example from the functional data perspective but serves as a toy example.
Then, the optimal solution $\tilde\beta$ can be given explicitly as
\[
\tilde\beta(t) = \left\{\begin{array}{ll} \delta(t)-\lambda & \mbox{ if } \delta(t)>\lambda \,,\\
0 & \mbox{ if } \delta(t) \in [-\lambda, \lambda]\,, \\
\delta(t) + \lambda & \mbox{ if } \delta(t) < -\lambda \,.\end{array}\right.
\]
Since $\delta$ is continuous, the set $\{t\in \cI: \tilde\beta(t) = 0\}$ will be a union of intervals and then $\tilde\beta$ joins the zero intervals continuously at the boundary. Hence the solution exhibits a thresholding behaviour, similar to the sparse estimators in the finite dimensional case, but on the continuous domain, thus justifying the notion of functional sparsity in the sense of domain selection. When $\eta=0$, the optimization occurs in $L^2(\cI)$, so the boundary points of the zero intervals are not necessarily differentiable. 
For the general case, it is difficult to express the solution in an analytical form but the same argument holds and when $\eta>0$ as in (\ref{e:sol1}) and (\ref{e:sol-alpha}), the function values at zero intervals and non-zero intervals are joined smoothly. 

\begin{figure}[tbp]
    \caption{Illustration of the roles of the tuning parameters $\eta$ and $\lambda$. Each panel plots the true $\beta_0(t)$, as well as $\tilde\beta(t)$ with a varying tuning parameter. } \label{fig:tuning}
    \begin{subfigure}[b]{.5\textwidth}
        \centering
        \includegraphics[width = \textwidth, height = 1.5in]{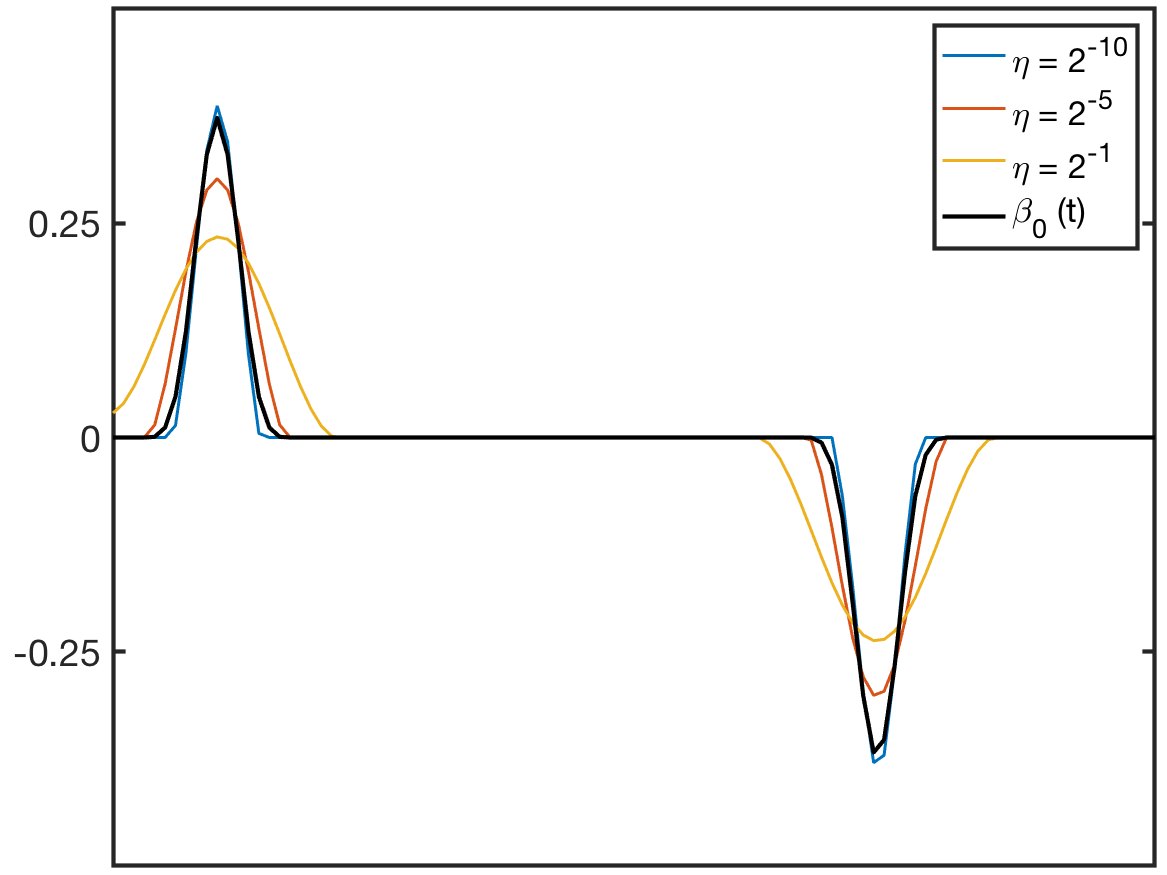}
        \caption{Setting 3, $\lambda = 2^{-7}$}
    \end{subfigure}%
            \begin{subfigure}[b]{.5\textwidth}
        \centering
        \includegraphics[width = \textwidth, height = 1.5in]{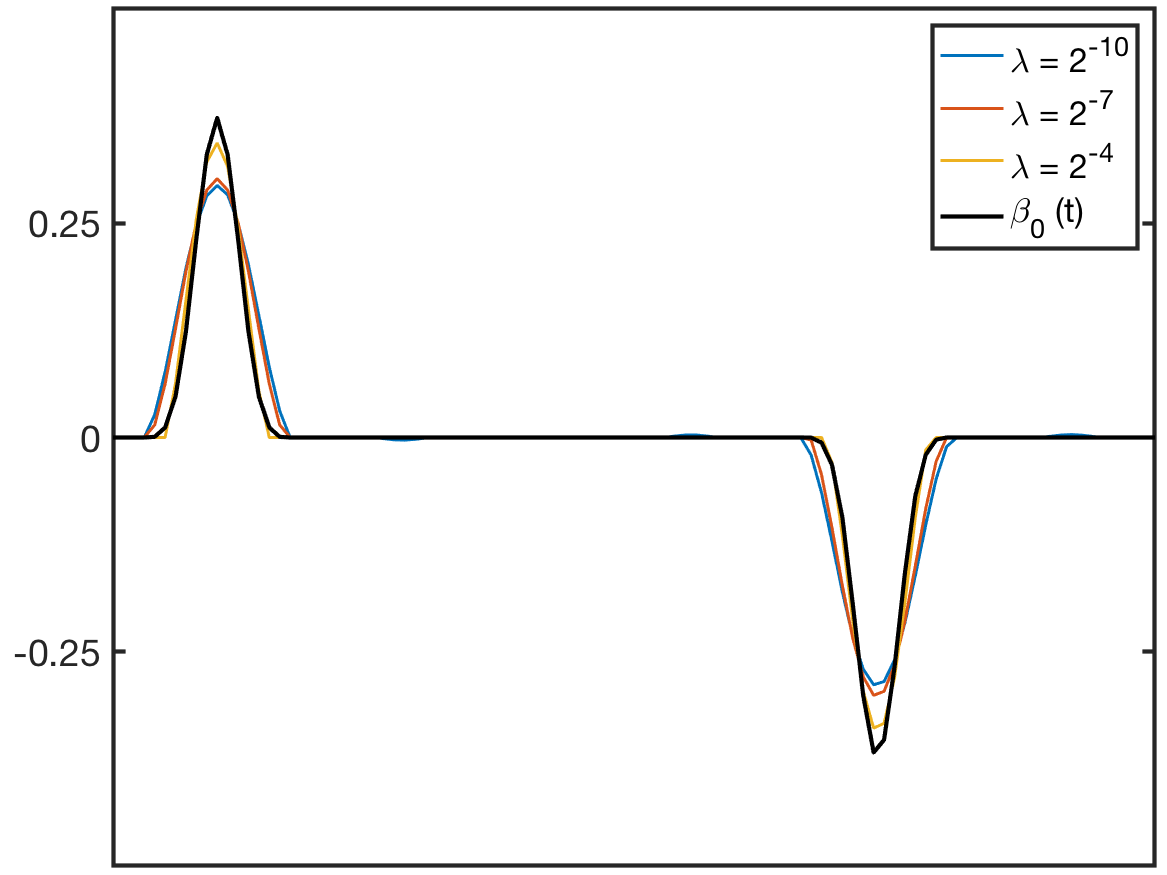}\\
        \caption{Setting 3, $\eta = 2^{-5}$}
    \end{subfigure}%
    
    \begin{subfigure}[b]{.5\textwidth}
        \centering
        \includegraphics[width = \textwidth, height =1.5in]{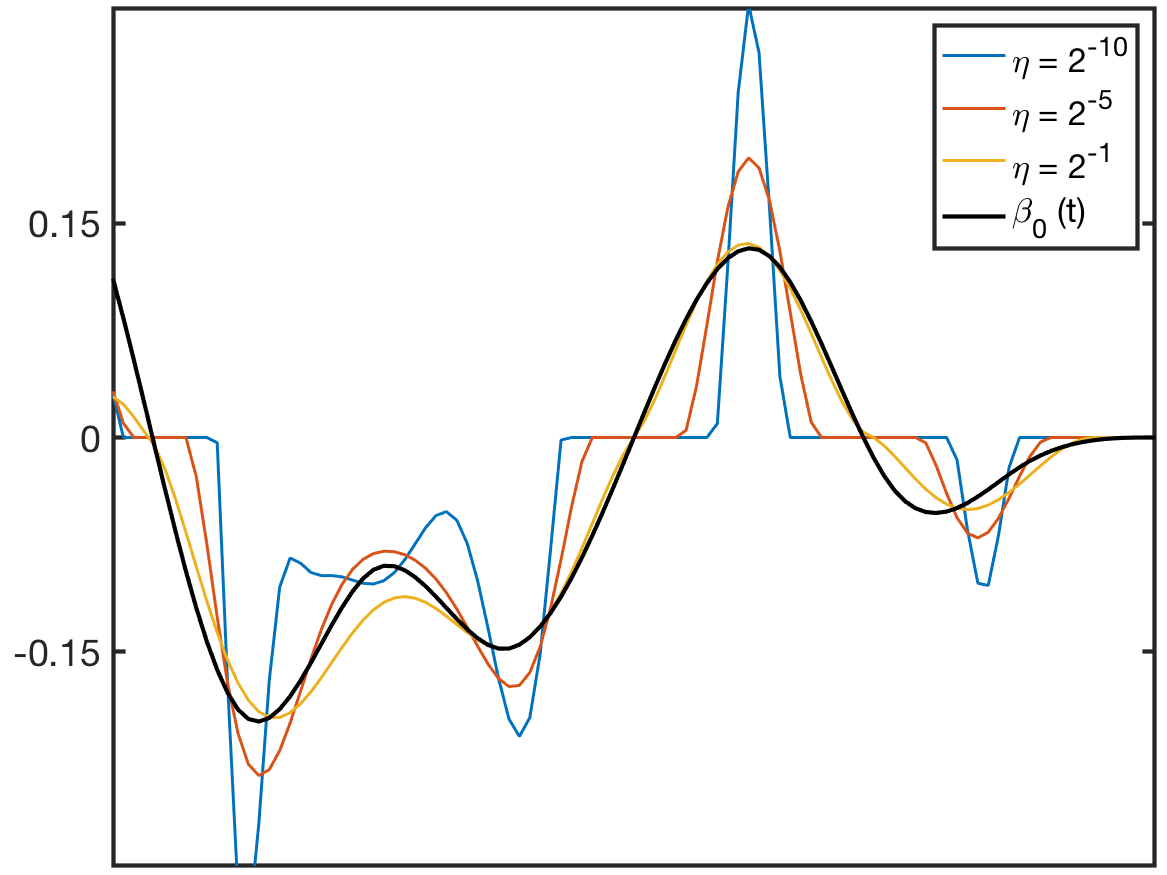}
        \caption{Setting 5, $\lambda = 2^{-7}$}
    \end{subfigure}
        \begin{subfigure}[b]{.5\textwidth}
        \centering
        \includegraphics[width = \textwidth, height = 1.5in]{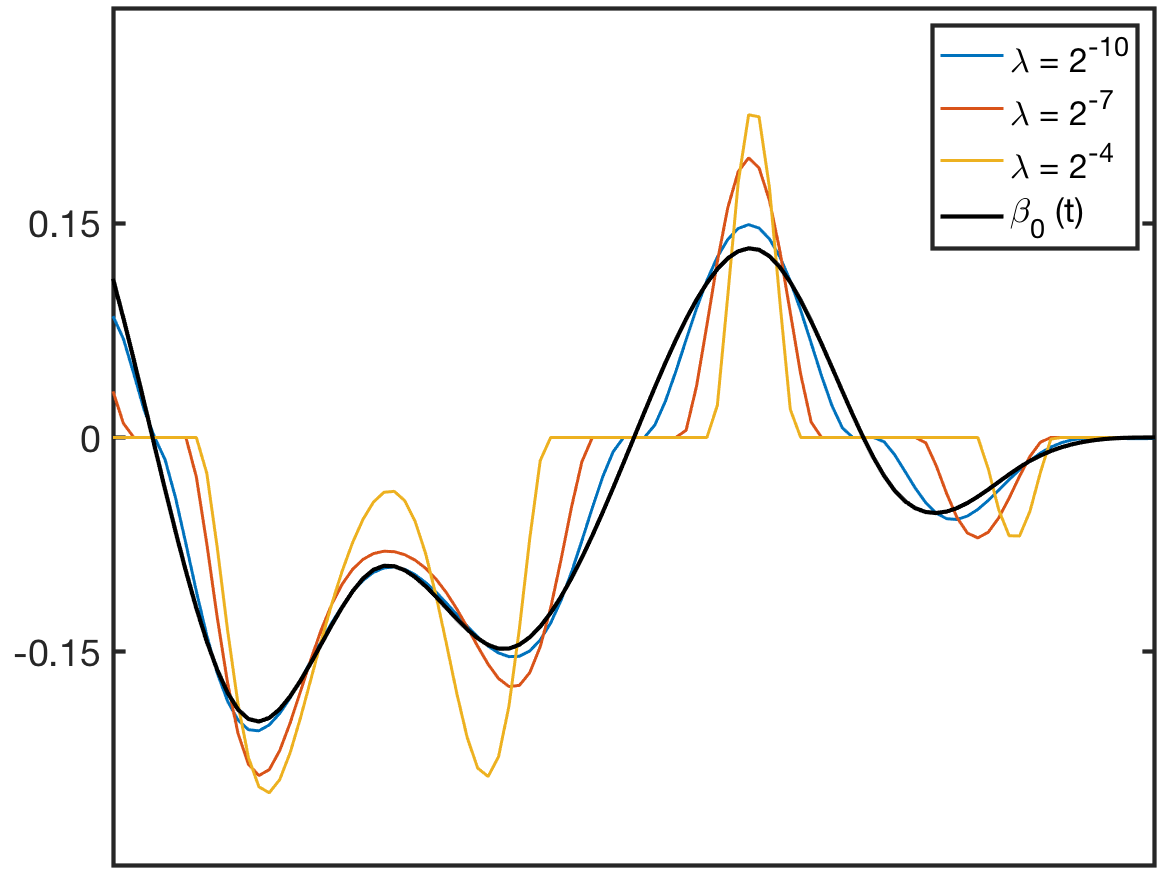}
        \caption{Setting 5, $\eta = 2^{-5}$}
    \end{subfigure}%
    \end{figure}

Moreover, suppose that $\beta_c \in H_0^1(\cI)  = \{u \in H^1(\cI): u \mbox{ is zero at the boundary of } \cI\}$. Then, (\ref{e:sol1}) leads to
\begin{equation}\label{e:sol2}
\int_\cI (\Gamma\tilde\beta - \delta + \alpha)\beta_c + \int_\cI (\eta\tilde\beta^\prime) \beta_c^\prime = 0 \,,\quad\mbox{ for all }\quad \beta_c \in H_0^1(\cI) \,.
\end{equation}
This implies \citep[e.g.,][]{Brezis2011} that $\tilde\beta^\prime$, the derivative of the solution, is also absolutely continuous and satisfies 
\begin{equation}\label{e:sol3}
\eta\tilde\beta^{\prime\prime} = \Gamma\tilde\beta - \delta + \alpha\,, \quad a.e.\,,
\end{equation} 
which, together with (\ref{e:sol-alpha}), gives an additional necessary condition for the solution.  

The following observation helps us to determine the upper bound of the tuning parameter $\lambda$. Suppose that $\delta \in L^\infty(\cI)$. Then, we have $\langle \delta,u\rangle \leq \|\delta\|_\infty \|u\|_1$. In addition, it can be shown that the solution $\tilde\beta$ satisfies 
\[
\langle\Gamma\tilde\beta, \tilde\beta\rangle + \eta \langle \tilde{\beta}^\prime, \tilde{\beta}^\prime\rangle +  \lambda\|\tilde\beta\|_1 = \langle\delta, \tilde\beta\rangle \leq \|\delta\|_\infty\|\tilde\beta\|_1 \,.
\]
It follows that 
\[
\langle\Gamma \tilde\beta, \tilde\beta\rangle + \eta \langle \tilde{\beta}^\prime, \tilde{\beta}^\prime\rangle + (\lambda - \|\delta\|_\infty) \|\tilde\beta\|_1 \leq 0 \,.
\]
Since the first two terms are non-negative, if $\|\delta\|_\infty \leq \lambda$, then $\tilde\beta=0$. This gives a range of $\lambda$ values for a non-zero solution and this depends on the magnitude of $\delta$.  Figure \ref{fig:tuning} illustrates the effect of the tuning parameters $\eta$ and $\lambda$ on $\tilde\beta$, under two different scenarios for $\beta_0$: sparse and non-sparse. The curves in the top and bottom panels respectively correspond to the simulation settings 3 and 5 in Section \ref{sec:numeric}.
We observe that for fixed $\lambda$, larger $\eta$ gives smoother estimates and for fixed $\eta$, larger $\lambda$ gives a more sparse solution.

\medskip
Suppose that $\mbox{Ker}(\Gamma) = \{0\}$. If $\|\Gamma^{-1}\delta\|_2<\infty$, then $\beta_0 = \Gamma^{-1}\delta\in L_2(\cI)$ is the unique solution to (\ref{e:lda}). Denote by $\tilde\beta$ the minimizer of $J$ in (\ref{e:J1}). The following proposition shows that the regularized solution $\tilde\beta$ approximates the underlying model solution to (\ref{e:lda}) whether the latter is finite or not.   
We write $\tilde\lambda = (\lambda, \eta)$, and $\tilde\lambda\rightarrow 0$ means that $\lambda\rightarrow 0$ and $\eta \rightarrow 0$. 
\begin{proposition} \label{prop:conv-beta} 
Fix any $u\in L_2(\cI)$. If $\|\Gamma^{-1}\delta\|_2< \infty$, then 
$\langle \tilde\beta, u \rangle \rightarrow \langle \beta_0, u\rangle$ as $\tilde\lambda \rightarrow 0$.
If $\|\Gamma^{-1}\delta\|_2=\infty$, 
then $\|\tilde\beta\| \rightarrow \infty$, as $\tilde\lambda \rightarrow 0$. 
\end{proposition}

Using Proposition~\ref{prop:conv-beta} together with (\ref{e:error-gauss}), we can study the misclassification error of the regularized classifier. We show below that our regularized classifier mimics the behaviour of the optimal classifier in terms of the misclassification rate.

\begin{proposition}\label{prop:error} 
Assume that $X$ is Gaussian with covariance operator $\Gamma$. The misclassification probability of the regularized classifier $\tilde\beta$, denoted by $err_X(\tilde\beta)$,
converges to $1 - \Phi(\|\Gamma^{-1/2}\delta\|_2/2)$ 
as $\tilde\lambda \rightarrow 0$. 
\end{proposition}
 
\begin{remark} It may be suspected that the two classes have different covariance functions, say $\Gamma_0$ and $\Gamma_1$, in which case a full generalization of our linear approach is not trivial. 
For example, a direct generalization of (\ref{e:fisher}) replaces the denominator by $\langle \tilde\Gamma\beta, \beta \rangle$ where $\tilde\Gamma = \pi_0\Gamma_0 + \pi_1\Gamma_1$, the pooled covariance function. Then the same formulation as (\ref{e:J1}), with $\Gamma$ replaced by $\tilde\Gamma$, follows.  
Alternatively, we can consider the two-dimensional projection approach for quadratic discriminant analysis proposed by \cite{gaynanova2019sparse} in the multivariate setting. In our functional framework, we would optimize the following two objective functions
\[
J_i(\beta) = \frac{1}{2}\langle(\Gamma_i \beta, \beta \rangle -\langle \delta, \beta \rangle + \lambda_i \|\beta\|_1 + \frac{\eta_i}{2}\|\beta'\|_2^2,
\]
for $i=0,1$. The classification will be then made based on the two linear maps with the respective solutions. An advantage of this method over standard quadratic discriminant analysis is that the computational complexity is essentially the same as the homoscedastic case being considered here. 
In either case, the theoretical guarantee on the domain selection and classification accuracy need to be carefully investigated in the functional context.  
\end{remark} 
 
\subsection{Properties of the regularized classifier in the sample model} \label{sec:sol-sample}

The development in the previous section justifies the use of the $L^1$ penalty in our formulation, by ensuring that we have a well-defined regularization problem (Proposition~\ref{prop:existence}) with the desired sparsity in the solution (Proposition~\ref{prop:solution}). 
We can show that these properties also hold for the empirical estimator, denoted by $\hat\beta$, the minimizer of $J_n$ defined in (\ref{e:Jn}). 

In the sample case, our formulation of $J_n$ is based on two estimators $\delta_n$ and $\Gamma_n$. 
Provided that the sample covariance operator $\Gamma_n$ is non-negative definite and $\delta_n \in L^2(\cI)$, it can be shown that Propositions~\ref{prop:existence} and \ref{prop:solution} hold for $J_n$ and $\hat\beta$. In fact, the proofs are essentially the same, thus we omit the formal statements of the Corollaries here.

In order to study the limit of the empirical estimator $\hat\beta$ and the corresponding misclassification probability, we make the following assumptions. Assumption~\ref{A:delGam}  says that some consistent estimators are available for $\delta_n$ and $\Gamma_n$  (or $\gamma_n$) and Assumption~\ref{A:lam-n} assumes that the tuning parameters decrease not too rapidly as the sample size gets large.

\begin{assumption}\label{A:delGam} Let $(\delta_n, \gamma_n)$ be consistent estimators such that
\[
\|\gamma_n - \gamma\|_\infty = O_p(a_n), \qquad  \|\delta_n- \delta\|_\infty = O_p(b_n)\,,
\]
for some sequences $a_n, b_n$ converging to 0. 
\end{assumption}
\begin{assumption}\label{A:lam-n} 
Let $\lambda=\lambda_n \rightarrow 0, \eta=\eta_n \rightarrow 0$ with $\lambda_n/\eta_n = O(1), a_n/\eta_n = o(1)$ and $b_n/\eta_n = o(1)$.
\end{assumption}
For sufficiently dense data as we consider here, the standard sample mean and sample covariance function can be used with a root-$n$ convergence.
Alternatively, any preferred smoothing methods could be employed, then the convergence rates are given as a function of smoothing parameters used, for example, see \citet{Yao2005Jun, LiHsing2012} for local linear smoothers. Normally the error for covariance estimation is larger than that for the mean estimation ($a_n \geq b_n$). \citet{ZhangWang2016} generalize those results under a general design scheme and show that a root-$n$ rate can be achieved for dense and ultra-dense data, in which case the uniform convergence rates can be $a_n=b_n = (\log n/n)^{1/2}$. 

\begin{proposition}\label{prop:conv-beta2} Fix any $u\in L_2(\cI)$. Suppose that Assumptions~\ref{A:X}-\ref{A:lam-n} hold. 
If $\|\Gamma^{-1}\delta\|_2<\infty$, then 
$\langle\hat\beta, u\rangle \rightarrow \langle \beta_0, u\rangle$ in probability as $n \rightarrow\infty$. 
On the other hand, if $\|\Gamma^{-1}\delta\|_2=\infty$, 
then $\|\hat\beta\| \rightarrow \infty$ in probability as $n \rightarrow \infty$. 
\end{proposition}

\begin{proposition}\label{prop:error2}  
Assume that $X$ is Gaussian with covariance operator $\Gamma$. Under Assumptions~~\ref{A:X}-\ref{A:lam-n}, 
the misclassification probability of the regularized classifier $\hat\beta$, denoted by $err_X(\hat\beta)$, converges to $1 - \Phi(\|\Gamma^{-1}\delta\|_2/2)$ in probability as $n \rightarrow\infty$. 
\end{proposition}

\subsection{Numerical algorithms} 

A fundamental strategy to solve infinite-dimensional optimization problems could be said to be either to discretize-and-optimize or to optimize-and-discretize \citep{Reyes2015}. For example, \citet{Glowinski1984} analysed numerical methods based on the former approaches using piecewise linear and quadratic approximations on $H^1(\cI)$, while \citet{Reyes2015} includes examples of the latter. Generally, the former is easier, while the latter may give a more elegant solution. 
According to Proposition~\ref{prop:solution}, the second approach requires solving equations (\ref{e:sol1}), or (\ref{e:sol3}), and (\ref{e:sol-alpha}) simultaneously. Based on these equations, Newton-like update steps for both $\tilde\beta$ and $\alpha$ could be derived. However, as this involves further development of the testable conditions at the iteration to be adapted to a proper function space \citep{Reyes2015}, we leave it for future work. 

Here we have used piecewise linear approximations in our implementation with equidistant grid points. We use estimates from the method 
by \citet{Yao2005Jun} for $\Gamma_n$ and $\delta_n$ and evaluate the derivative by the finite-difference approximation. 
This strategy, combined with the rather simple form of the optimization problem, leads to a lasso-type problem, for which many efficient optimization algorithms are available. 
We implement the coordinate descent algorithm by \citet{Fu2012Feb} to solve our discretized problem. 
\subsection{Implications in functional linear regression}

Although we started with a classification problem, our regularization method with an $L^1$-type penalty can be motivated from a regression point of view.
Consider a functional regression problem with scalar response $Y$ and functional predictor $X \in H = L^2(\cI)$. For simplicity, assume that $\E(Y) = 0$ and $\E(X) = 0$ and consider a linear regression given by
\[
Y = \int_{\cI} \beta(t)X(t)\,\mbox{d}t + \epsilon \,,
\]
where $\E(\epsilon) = 0$ and $var(\epsilon) = \sigma^2$. 
As before, the covariance operator of $X$ is denoted by $\Gamma$. In addition, let $\Lambda:L^2(\cI) \rightarrow \mathbb{R}$ denote the covariance operator between $X$ and $Y$, defined as 
\[
\Lambda(u) = \int_{\cI} \E\{ X(s)Y\} u(s)\,\mbox{d}s \,.
\]
Then, the population least squares criterion can be expressed as
\[
E\{(Y - \langle\beta, X\rangle)^2\} = \langle\Gamma \beta,\beta\rangle - 2 \Lambda(\beta) + const \,,
\]
from which it follows that if $\tilde\beta$ is the minimizer if and only if it satisfies
\begin{equation} \label{e:sol-fln}
\langle\Gamma\tilde\beta, u\rangle = \Lambda (u) \qquad \mbox { for all } u \in H \,.
\end{equation}
To be consistent with the earlier notation, write $\Lambda (u) = \langle \Delta, u\rangle$ where $\Delta(\cdot) = E\{X(\cdot)Y\}$. Then, (\ref{e:sol-fln}) can be expressed as
\[
\langle \Gamma\tilde\beta, u \rangle = \langle \Delta, u\rangle \qquad \mbox{ for all }  u \in H \,.
\]
Note that when $Y$ is binary with values in $\{-1, 1\}$, $\Delta$ is equal to $\delta$ in (\ref{e:lda}), the difference in the group means in $X$ between two classes. 

The existence of the solution requires the so-called Picard condition \citep{CardotFerratySarda2003, Hsing:Eubank:2015} 
\[
\sum_{j=1}^\infty \frac{\langle E\{X(\cdot)Y\}, \phi_j\rangle^2}{\theta_j^2}   = \sum_{j=1}^\infty \frac{\langle\Delta, \phi_j\rangle^2}{\theta_j^2} < \infty \,,
\]
which is equivalent to $\| \Gamma^{-1}\delta\|_2 < \infty$ in the case of classification problems (c.f., Section~\ref{sec:optbeta}).

The above discussion shows that even though the motivation differs, the underlying problem for functional linear regression is equivalent to that of linear classification. 
Hence, our methodology is equally applicable in a regression setting and offers a regularization method to estimate the coefficient function $\beta$. 
Given a sample of size $n$ of $(Y_i, X_i), i=1,\ldots,n$, denote the sample version of the operators of $\Gamma$ and $\Lambda$ by $\Gamma_n$ and $\Lambda_n$, respectively, defined as
\[
\Gamma_n (u)(t) = \frac{1}{n}\sum_{i=1}^n \langle X_i, u\rangle X_i(t) \,,\quad \Lambda_n (u) =  \frac{1}{n}\sum_{i=1}^n \langle X_i, u\rangle Y_i  = \langle \Delta_n, u\rangle  \qquad u \in L^2(\cI) \,.
\]
where $\Delta_n(\cdot) = (1/n)\sum_{i=1}^n X_i(\cdot) Y_i$. Assuming that the true coefficient function $\beta$ is a smooth function, \citet{CardotFerratySarda2003} proposed a generalization of a ridge regression with an $L^2$ derivative penalty as
\[
\frac{1}{2}\langle \Gamma_n \beta_K, \beta_K \rangle - \langle \Delta_n, \beta_K\rangle + \frac{\eta}{2} \|\beta_K^{(m)}\|_2^2 \,,
\]
where $\beta_K = \sum_{j=1}^K c_k B_k$ with $B_k$ is a B-spline basis function and $\beta_K^{(m)}, m\geq 1$ is the $m$th derivative of $\beta_K$.
Our analysis suggests that adding an $L^1$ penalty would lead to a sparse estimator that consistently estimates the true function, whether or not the latter is sparse. 

\begin{remark}
Although the theoretical framework differs, our main consideration of interpretability is not limited to this type of problem. For example, semi-parametric problems of single-index or multiple-index regression or classification can be expressed as
\[
Y = F\left(\sum_k \langle X, \beta_k\rangle\right) + \epsilon
\]
with a known or unknown link function $F$.
Then an $L^1$-type functional regularization for $\beta_k$ can be developed with an appropriate loss function. We leave such a generalization to our future work.   
\end{remark}

\section{Numerical studies} \label{sec:numeric}

\subsection{Simulation models}

We use simulated curves under six different underlying population scenarios to  assess the performance of the proposed method and compare it with existing approaches. Settings 1 -- 5 represent scenarios where the underlying structure follows model (\ref{e:lda}) in Section \ref{sec:optbeta} and Setting 6 is borrowed from \citet{DelaigleHall2012}. The proposed approach, which we name SFLDA (Sparse Functional Linear Discriminant Analysis), is compared with eight different approaches: the non-sparse version of the proposed approach (FLDA) that sets the tuning parameter $\lambda = 0$ in (\ref{e:Jn}); ridge FLDA (RFLDA) proposed by \citet{kraus2018classification}; partial least squares (PLS) by \citet{DelaigleHall2012}; the three Bayes methods (B-Gauss, B-NPD, B-NPR) in \citet{DaiMuellerYao2017}; the quadratic discriminant analysis (QDA) by \citet{galeano2015mahalanobis}; functional logistic regression (Logistic) \citep{muller2005functional}.

The first four settings have a sparse underlying discriminant function. The sparse discriminant functions, $\beta_1(t)$ and $\beta_2(t)$, are generated with  $33$ cubic B-spline basis functions $B_{i,4} ~( i = 1, \dots ,33)$ with the knots $\xi_j = j/30, j=0,\dots,30$, where $\xi_0$ and $\xi_{30}$ are the boundary knots. 
We set $\beta_1(t) = 0.2 B_{5,4} (t) + 0.2 B_{28,4} (t) $ for Settings 1 and 2, and 
 $\beta_2(t) = 0.2 B_{5,4} (t) - 0.2 B_{24,4} (t) $ for Settings 3 and 4. For the non-sparse $\beta_3(t)$ for Setting 5, we use 8 cubic B-spline basis and set 
 \[
 \beta_3(t) = 0.1 B_{1,4}(t) - 0.3 B_{3,4}(t) - 0.2 B_{5,4}(t) + 0.2 B_{7,4}(t) - 0.1 B_{8,4}(t).
 \]
For the covariance function $\Gamma$, we used the Mat\'{e}rn covariance function:
\[\gamma(s,t)= \sigma^2\frac{ 2^{1-\nu}}{G(\nu)} \left\{(2 \nu)^{1/2} \frac{|s-t|}{\rho}\right\}^\nu  K_\nu \left\{(2 \nu)^{1/2} \frac{|s-t|}{\rho}\right\}, \]
where $G$ is the gamma function, $K_\nu$ is the modified Bessel function of the second kind, for which the R-function {\tt besselK} was used, and the parameters were set as $\sigma=1$, $\rho=0.2$, and $\nu=3$. We set the mean of the first group as
$\mu_0(t) = 5t+\sum_{j=1}^5 (c_j/j)\phi(t)$, where $\phi(t) = 2^{1/2}\sin(\pi j t)$, 
 $(c_1,\dots,c_5)= (2.19, -0.18, -0.19, -2.51, -0.56)$, and the second group mean as $\mu_1(t) = \mu_0(t) + \delta(t)$. Here the mean difference function $\delta(t)$ is determined by $\delta(t)  =  \int_\cI \Gamma(t,s) \beta_k(s) ds$, $k = 1,2,3$.  

The data were assumed to be available on a fine grid. We used $T = 100$ equispaced grid points on $[0,1]$. For grid points $t_j, j=1,\dots,T$ , let $\tilde\Gamma$ be the $T\times T$ matrix with $(i,j)$ entry $\Gamma(t_i,t_j)$, and
let $m_0$  and $m_1$ be the $T\times 1$ vectors with $i$th entry $\mu_0(t_i)$ and $\mu_1(t_i)$, respectively. Data were generated by
$x_0 = m_0 + \tilde\Gamma^{1/2} e_0 $ and $x_1 = m_1 + \tilde\Gamma^{1/2} e_1$
where $e_0$ and $e_1$ are the vectors of independent noise for each class, generated from either a standard normal for Settings 1, 3, and 5 or a $t$-distribution with $5$ degrees of freedom for Settings 2 and 4. 

As mentioned earlier, Setting 6 is borrowed from \citet{DelaigleHall2012}. 
Each curve from the $i$th group is generated as $x_i = \sum_{j=1}^{40}(j^{-1} Z_{ij} + \mu_{ij})\phi_j(t)$, where $Z_{ij}$ are centred exponential variables, $(\mu_{01}, \ldots,  \mu_{06}) = (0, -0.5, 1, -0.5, 1, -0.5)$, and $(\mu_{11}, \ldots, \mu_{16}) = (0, -0.75, 0.75, -0.15, 1.4, 0.1)$. 

We generated $n = 100$ curves for each group to train each classifier and independently generated 300 curves for each group to evaluate the misclassification error. The tuning parameters in each method are chosen via 5-fold cross-validation within the training data based on the misclassification error. 

\subsection{Results}

\begin{table}[tbp]
\caption{\label{tab:sim} Test error rates (\%) for simulated data based on 100 repetitions. Standard errors are in the parentheses.}
\begin{center}
{\small
\begin{tabular}{cccccccccc} 
Setting & SFLDA & FLDA & RFLDA & PLS & B-Gauss & B-NPD & B-NPR & Logistic & QDA \\ 
\multirow{ 2}{*}{1}&   33.3 &   33.4  &  33.6  &  33.7   & 33.4  &  34.6 &   35.1  &  34.7  &  34.2 \\
  & (0.22)  &  (0.22)  & (0.24)  &  (0.23)  &  (0.24)  &  (0.25)&    (0.25) &    (0.27)  &  (0.28) \\
 \multirow{ 2}{*}{2}&    33.5  &  33.5  &  33.8   & 33.7  &  33.6  &  34.7  &  35.4   & 34.9  &  34.2 \\
 &   (0.21) &    (0.20) &     (0.20)  &  (0.20) &    (0.20) &    (0.22) &    (0.24) &    (0.25) &    (0.22) \\ 
\multirow{ 2}{*}{3} &   37.2  &  36.9 &   37.5  &  37.1  &  37.2  &  37.8  &  39.0   & 38.5  &  38.0\\
 &  (0.23)  &  (0.20) &    (0.20)  &  (0.21) &     (0.22) &    (0.25) &     (0.29) &    (0.27) &    (0.25)\\ 
 \multirow{ 2}{*}{4}&   37.7  &  37.3  &  38.1  &  37.6  &  37.5  &  39.1  &  40.3   & 39.2 &   38.2\\
 &     (0.24) &    (0.22) &     (0.24) &     (0.23) &    (0.24) &    (0.27) &    (0.30) &     (0.28) &     (0.26)  \\
  \multirow{ 2}{*}{5} &   3.2 &   3.1  &  3.1  &  3.1  &  3.2  &  3.5  &  3.7   & 3.6 & 4.2 \\
 &   (0.08) &    (0.07)&    (0.07)&    (0.07)&    (0.08)&    (0.10)&    (0.10) &  (0.14) &    (0.08) \\
  \multirow{ 2}{*}{6} &  3.3  &  3.2  &  3.3  &  3.1   & 4.0  &  3.4 &   2.5  &  2.5  &  3.6 \\
 &     (0.08)&    (0.07)&    (0.07)&    (0.07)&    (0.08)&    (0.08)&    (0.10)&    (0.10)&    (0.14) \\
   \end{tabular}}
   \end{center}
\end{table}

\begin{center}
\begin{figure}[tbp]
    \caption{Discriminant functions from 100 repetitions of the simulation study estimated by SFLDA, FLDA, RFLDA, and PLS, for Settings 1, 3, and 5.  The thicker black line in each plot represents the median of the estimated curves, while the dot-dashed curve is the true discriminant function. }\label{fig:simulbeta}
    \begin{small}
        \begin{subfigure}[b]{.24\textwidth}
        \centering
        \includegraphics[width = \textwidth, height = 1in]{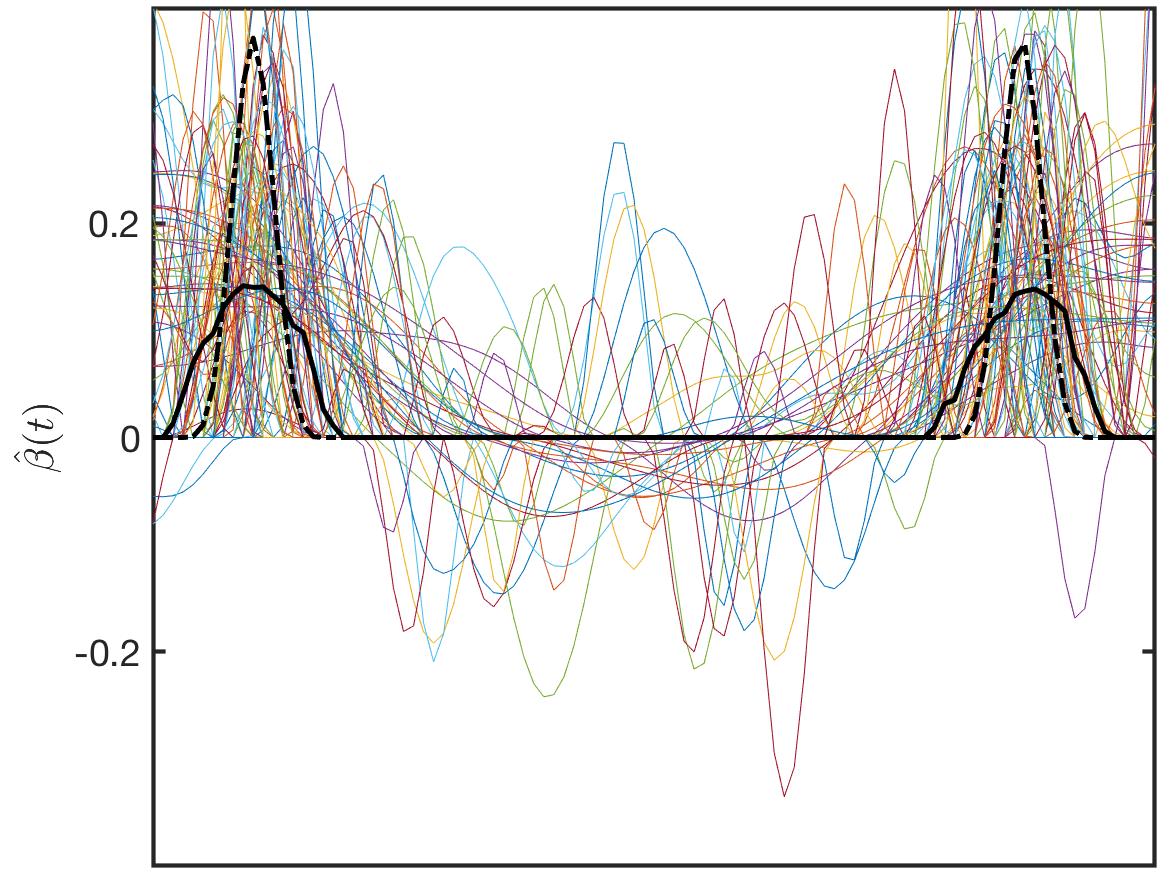}
        \caption*{SFLDA}
    \end{subfigure}%
            \begin{subfigure}[b]{.24\textwidth}
        \centering
        \includegraphics[width = \textwidth, height = 1in]{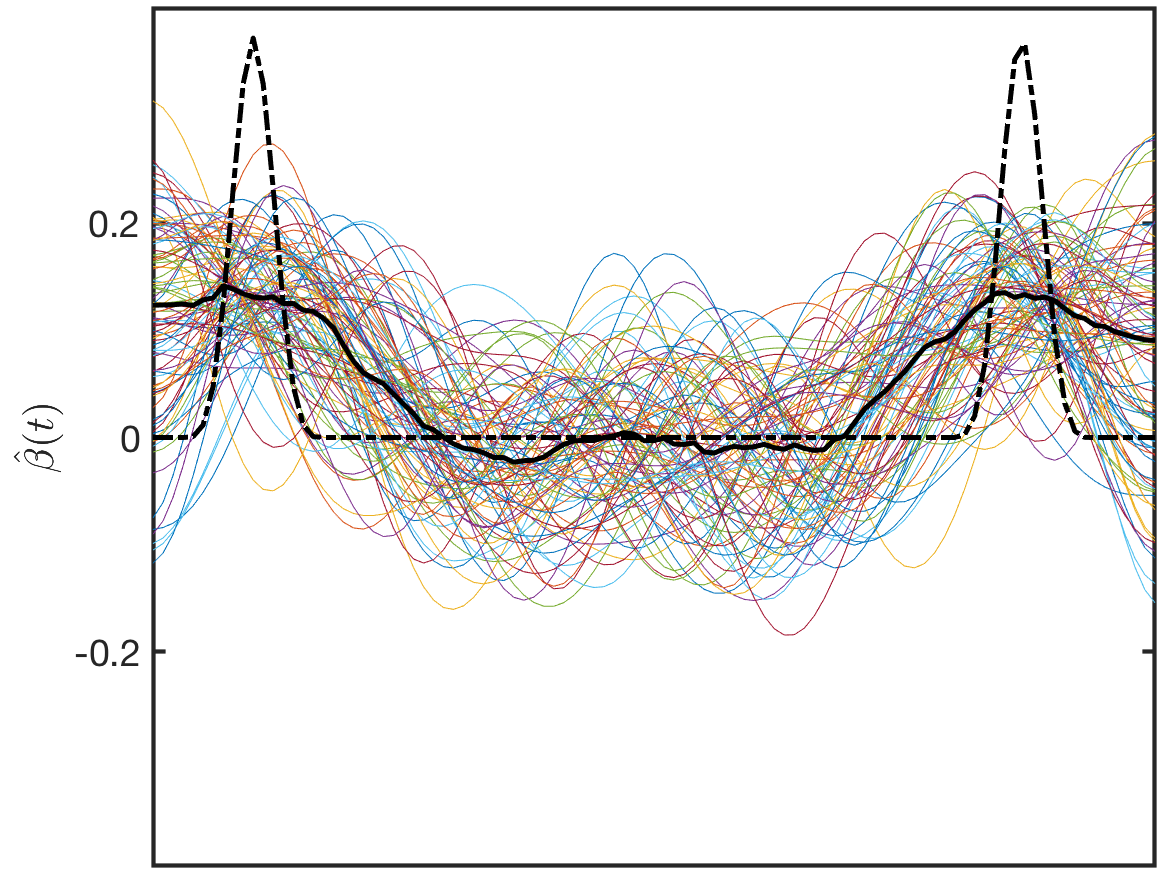} 
        \caption*{FLDA}
    \end{subfigure}%
           \begin{subfigure}[b]{.24\textwidth}
        \centering
        \includegraphics[width = \textwidth, height = 1in]{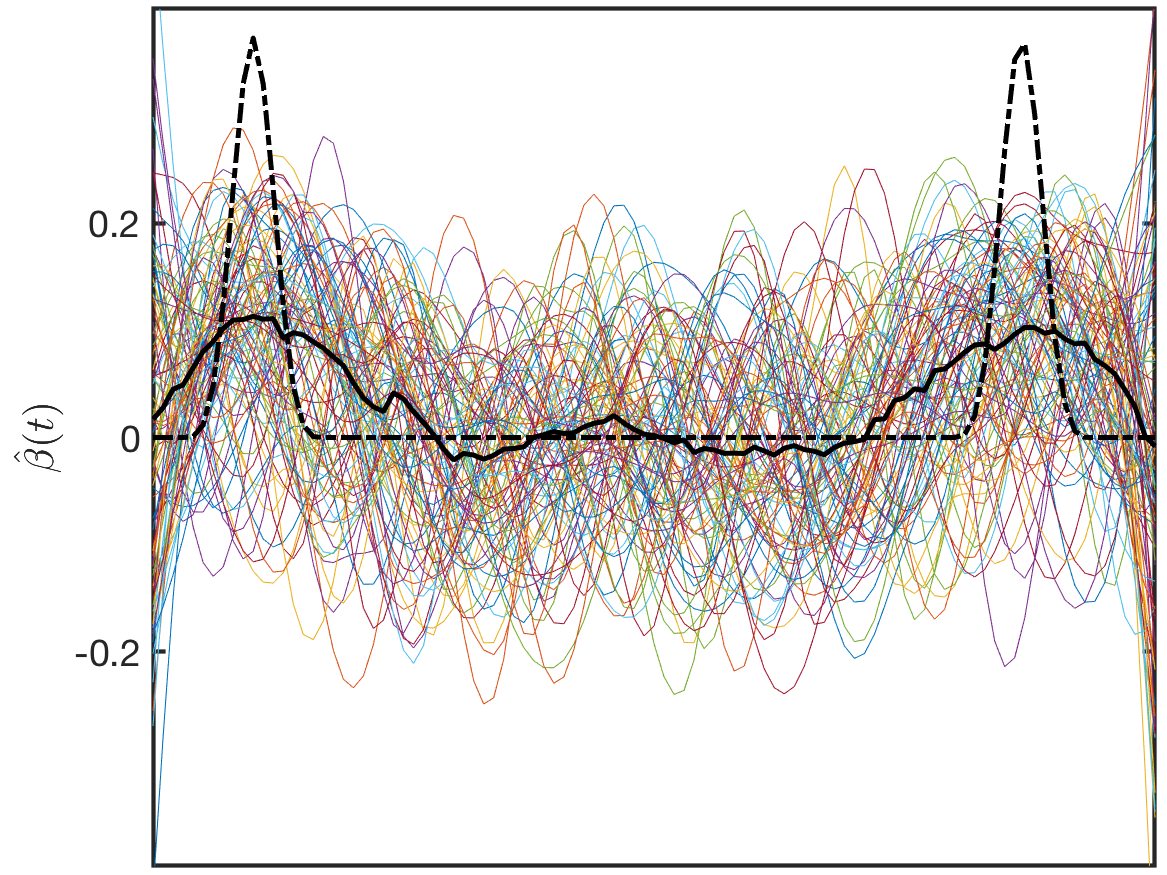}
        \caption*{RFLDA}
    \end{subfigure}%
           \begin{subfigure}[b]{.24\textwidth}
        \centering
        \includegraphics[width = \textwidth, height = 1in]{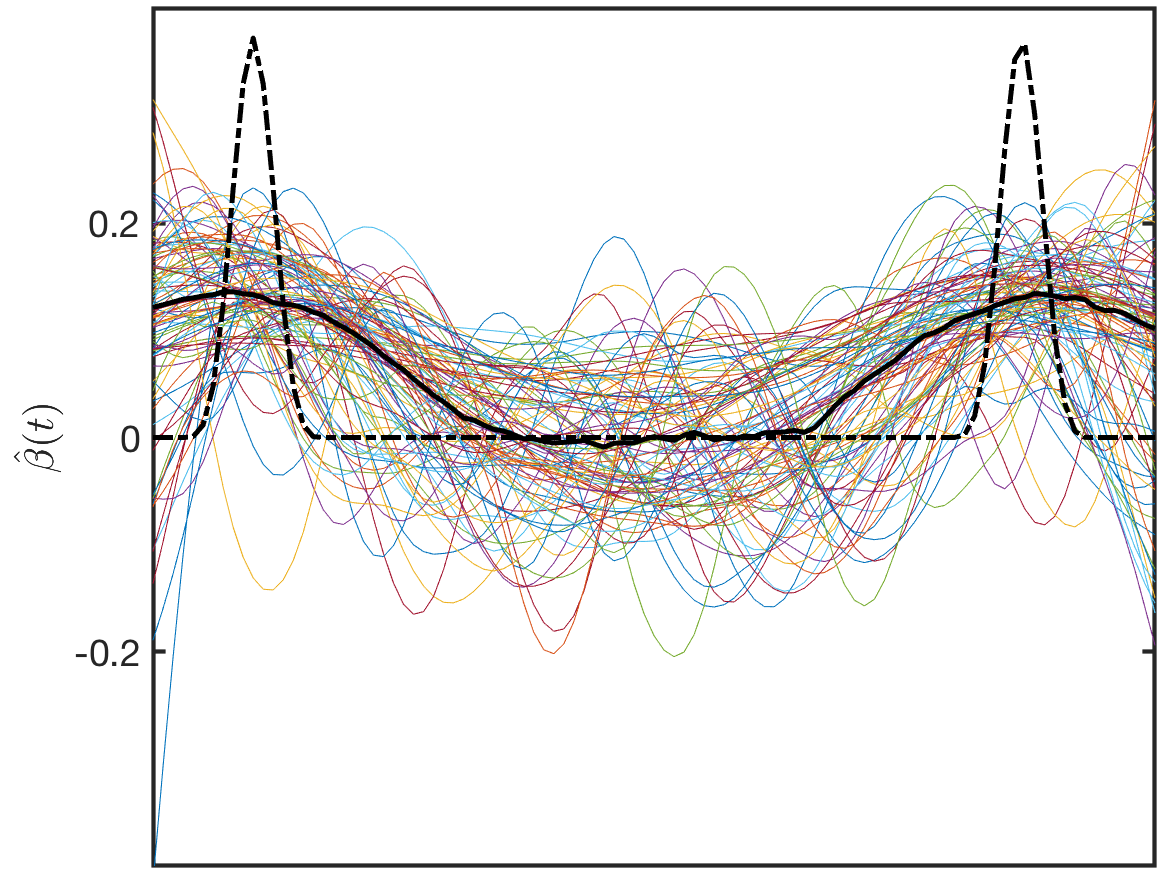}
        \caption*{PLS}
    \end{subfigure}\\
        \begin{subfigure}[b]{0.24\textwidth}
        \centering
        \includegraphics[width = \textwidth, height = 1in]{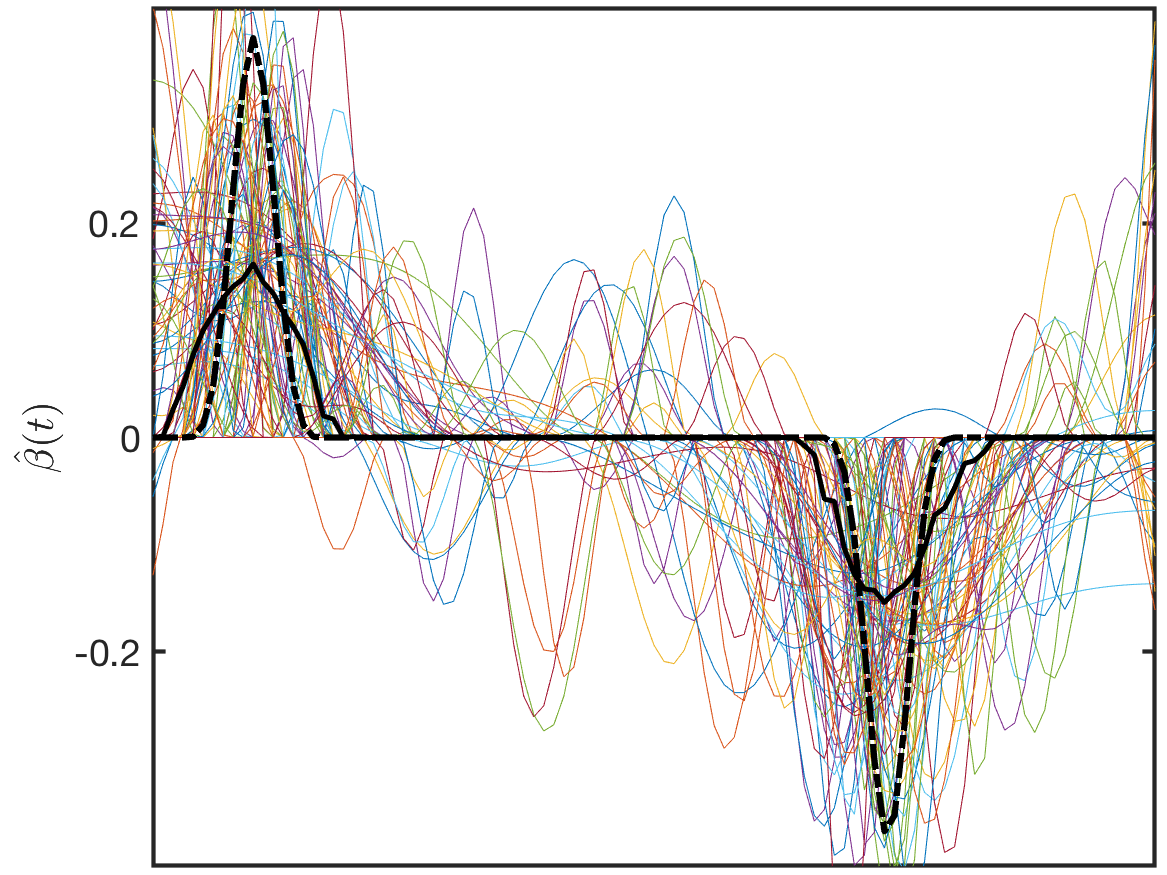}
        \caption*{SFLDA}
    \end{subfigure}%
           \begin{subfigure}[b]{0.24\textwidth}
        \centering
        \includegraphics[width = \textwidth, height = 1in]{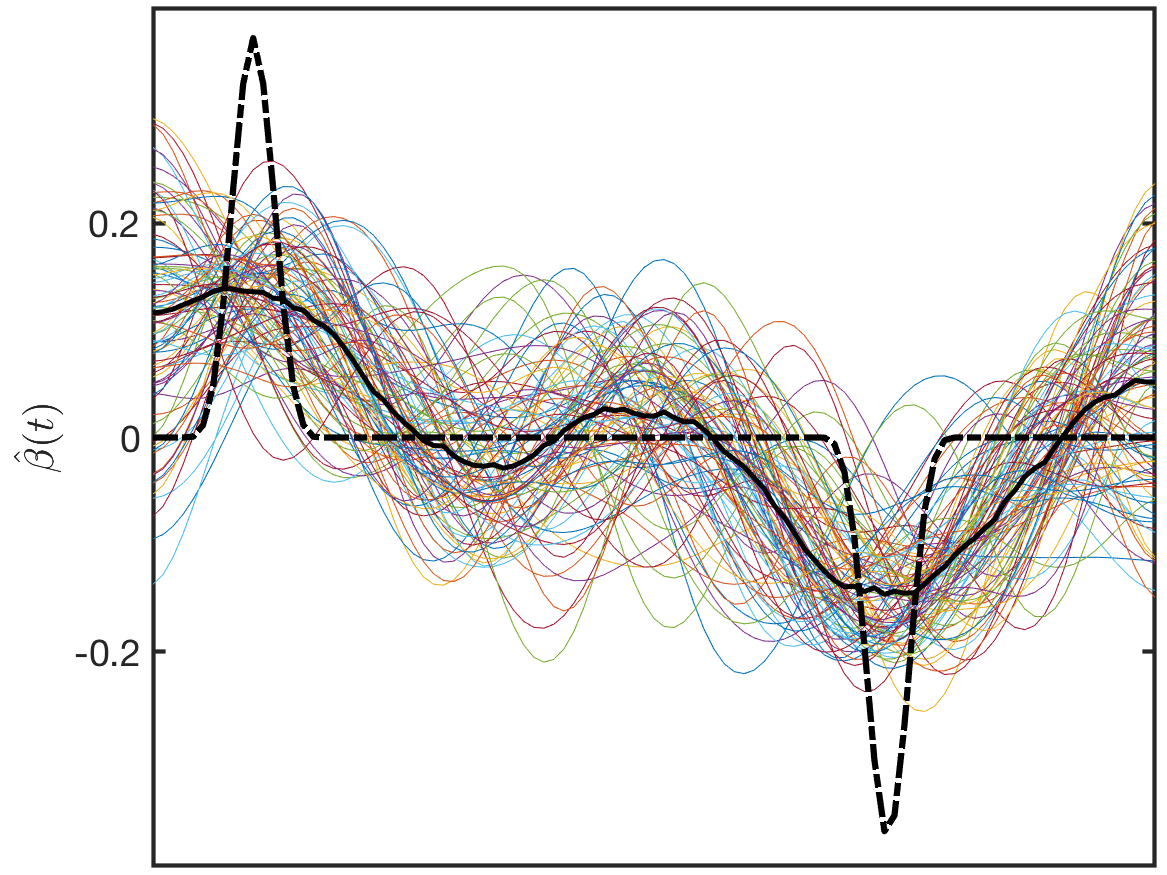}
        \caption*{FLDA}
    \end{subfigure}%
      \begin{subfigure}[b]{0.24\textwidth}
        \centering
        \includegraphics[width = \textwidth, height = 1in]{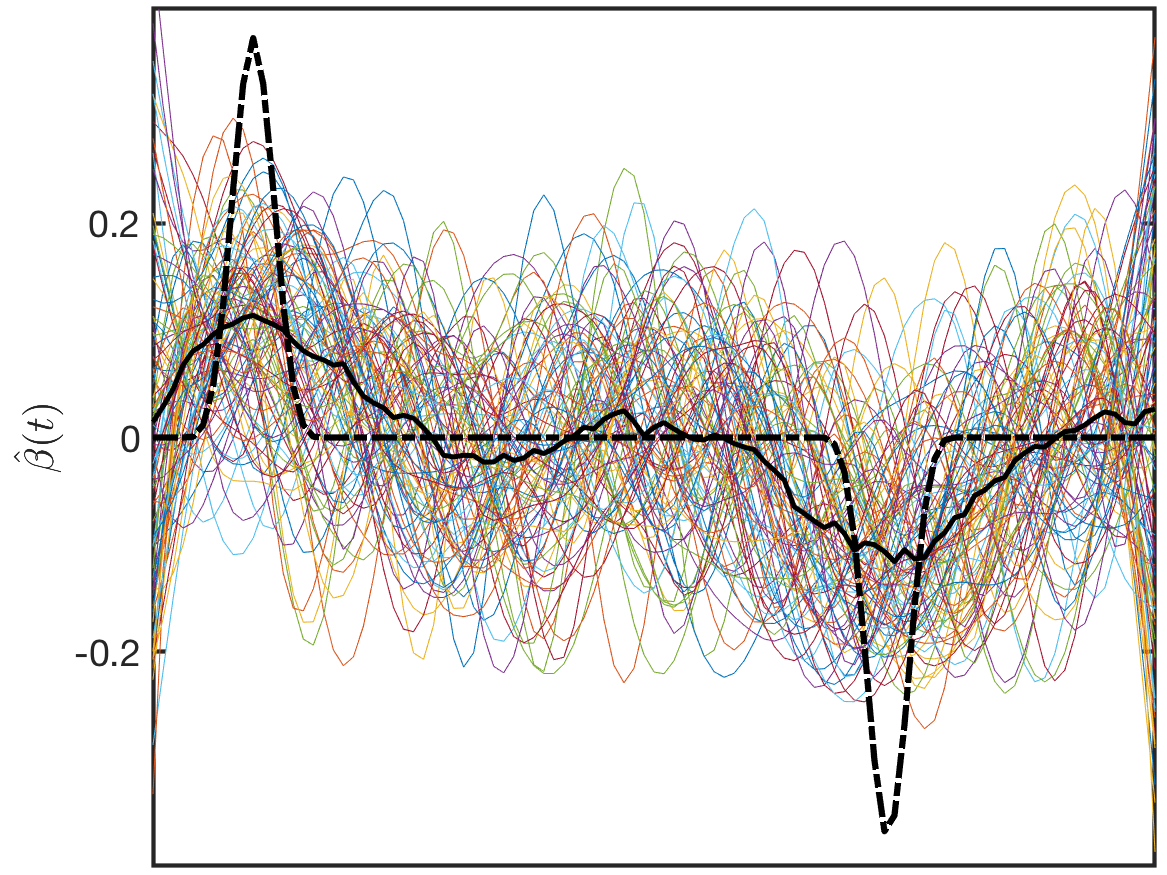}
        \caption*{RFLDA}
    \end{subfigure}%
       \begin{subfigure}[b]{0.24\textwidth}
        \centering
        \includegraphics[width = \textwidth, height = 1in]{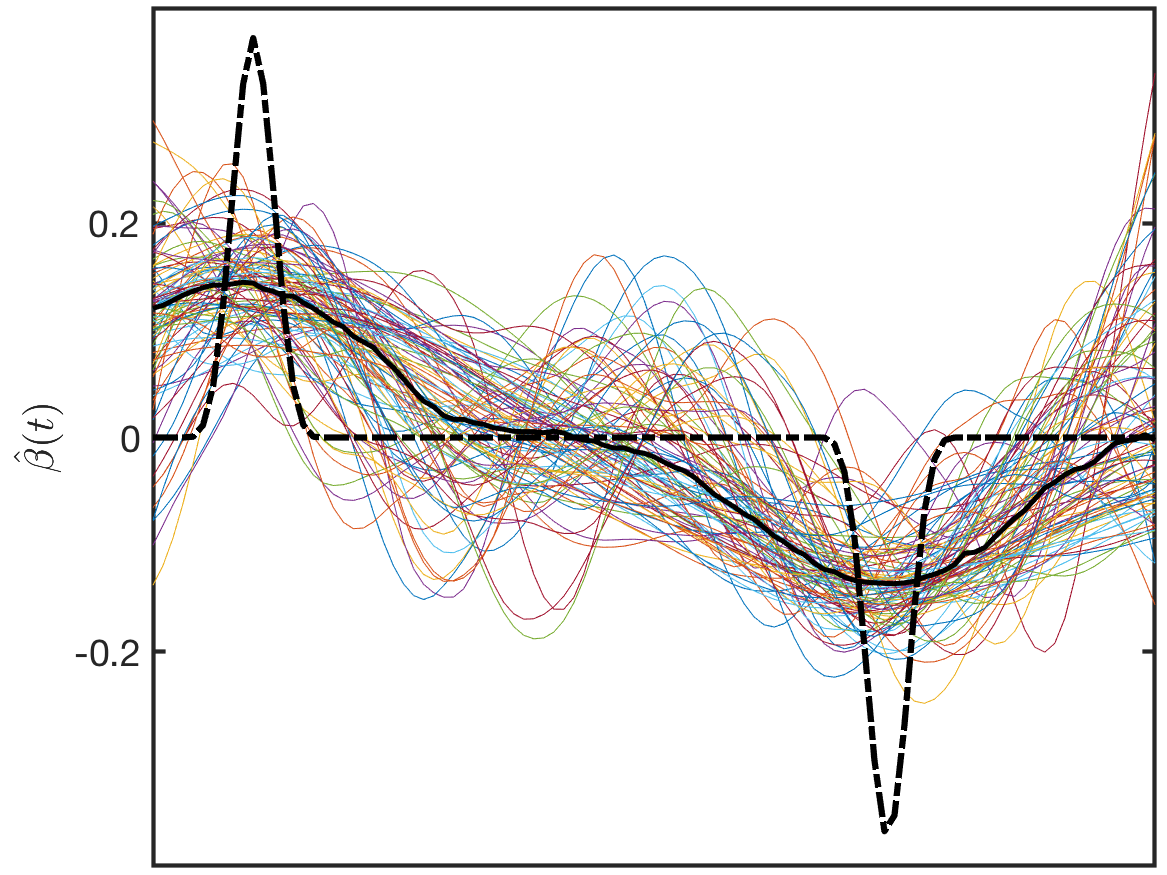}
        \caption*{PLS}
    \end{subfigure}\\
        \begin{subfigure}[b]{0.24\textwidth}
        \centering
        \includegraphics[width = \textwidth, height = 1in]{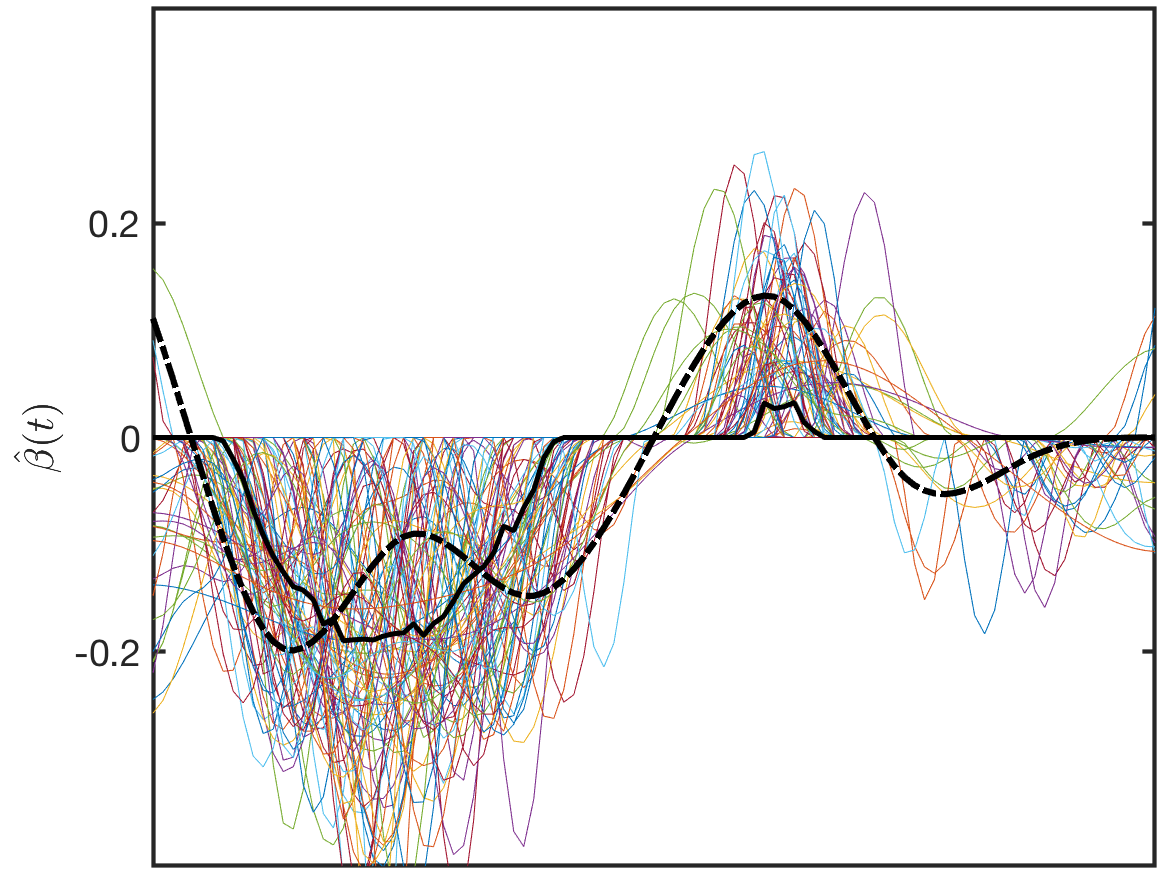}
        \caption*{SFLDA}
    \end{subfigure} 
         \begin{subfigure}[b]{0.24\textwidth}
        \centering
        \includegraphics[width = \textwidth, height = 1in]{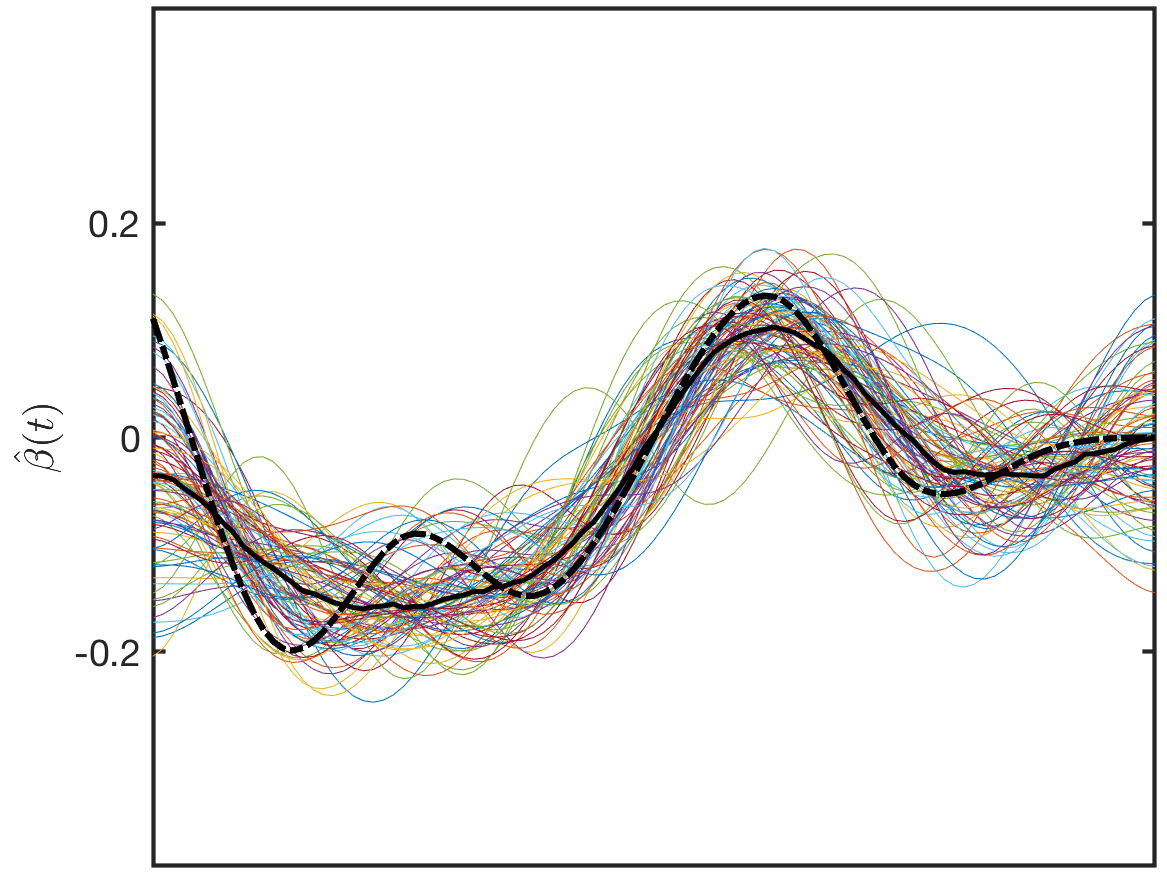}
        \caption*{FLDA}
    \end{subfigure} 
               \begin{subfigure}[b]{0.24\textwidth}
        \centering
        \includegraphics[width = \textwidth, height = 1in]{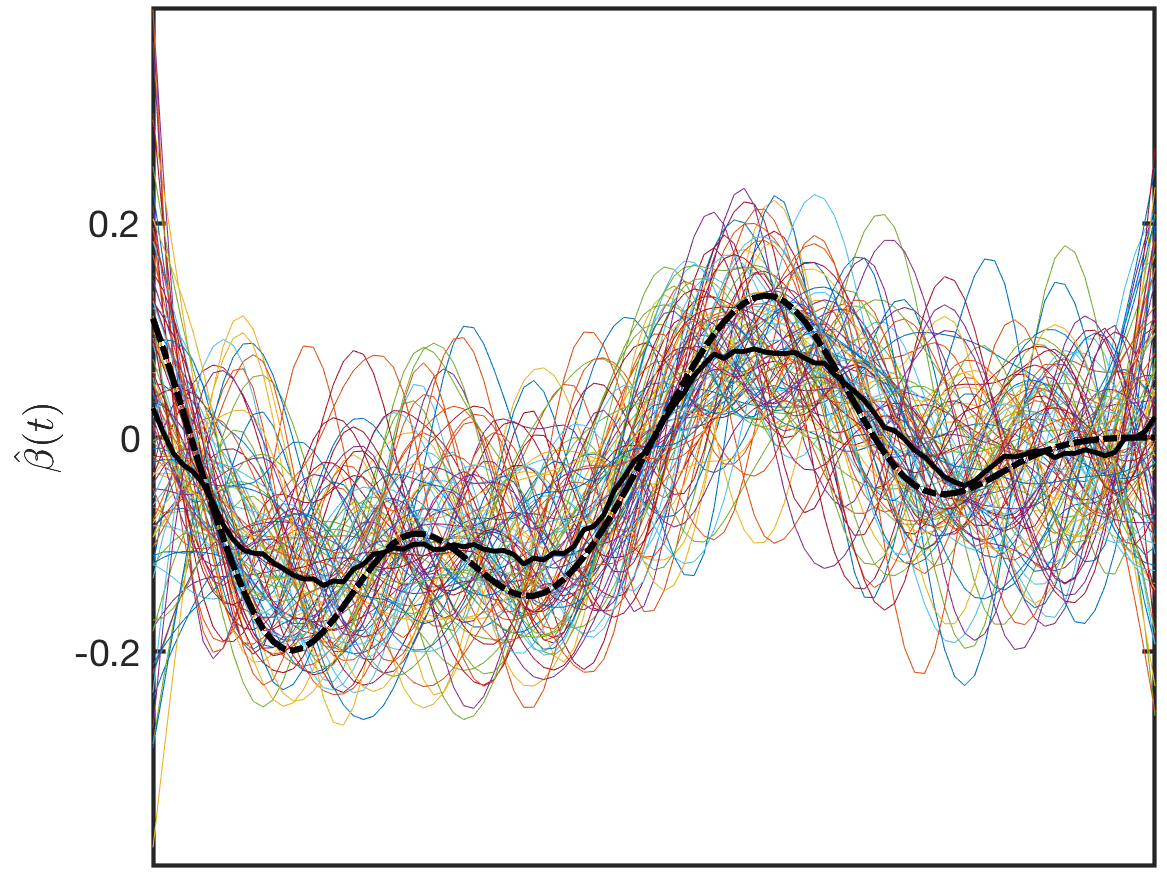}
        \caption*{RFLDA}
    \end{subfigure} 
          \begin{subfigure}[b]{0.24\textwidth}
        \centering
        \includegraphics[width = \textwidth, height = 1in]{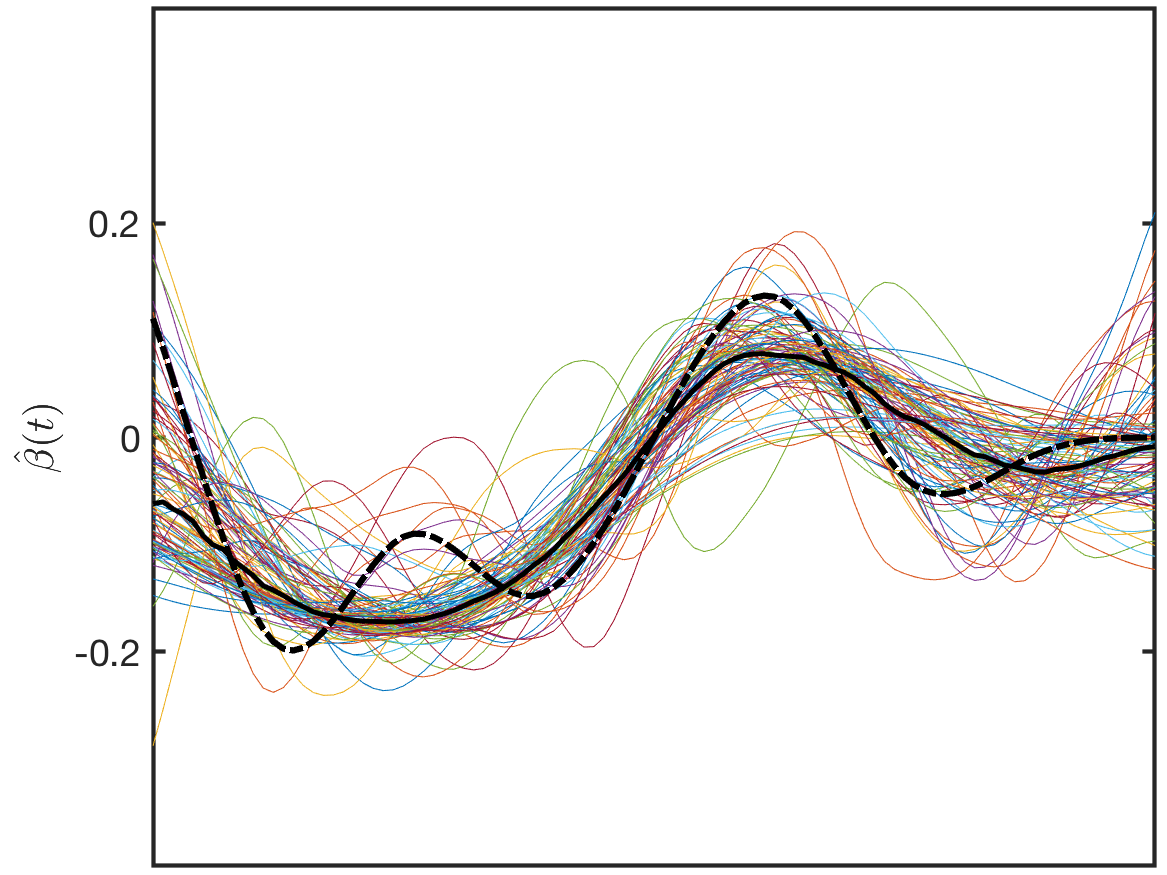}
        \caption*{PLS}
    \end{subfigure}%
    \end{small} 
\end{figure}
\end{center}

\begin{table}[tbp]
\caption{\label{tab:betanorm} Average norm difference ($\|\hat\beta - \beta_0\|_j$) between the estimated discriminant function and the true function. Numbers are multiplied by $100$.}
\begin{center}
{\small
\begin{tabular}{cccccc} 
Setting & $j$ & SFLDA & FLDA & RFLDA & PLS \\ 
\multirow{ 2}{*}{1}&  $1$ &  5.37 (0.19) &   7.97 (0.10) &  8.81 (0.13) &  8.20 (0.09) \\
  & $2$ & 9.64 (0.22) &  10.07 (0.11)& 11.29 (0.18)& 10.14 (0.10) \\
\multirow{ 2}{*}{3}&  $1$ &  5.84 (0.18)&   8.10 (0.11) &  9.00 (0.12) &  8.30 (0.11) \\
  & $2$ & 10.39 (0.22)&  10.40 (0.14) & 11.69 (0.18)& 10.26 (0.11)\\
\multirow{ 2}{*}{5}&  $1$ &  6.47 (0.14)&   4.15 (0.11)  &  5.78 (0.17) &  4.50  (0.10)\\
  & $2$ & 8.21 (0.18) &  5.25 (0.13)& 7.23 (0.20)& 5.64 (0.11) \\
   \end{tabular}}
   \end{center}
\end{table}
Table \ref{tab:sim} lists test errors based on 100 repetitions. The proposed approach is generally competitive in all settings, even though the differences among the methods do not stand out. A main benefit of the proposed sparse estimation can be seen in its ability to estimate zero regions in the discriminant function. 
Table \ref{tab:betanorm} displays the norm differences of the estimated discriminant function $\hat\beta(t)$ by the proposed method with sparse functional penalty, its non-sparse version, ridge regularization, and partial least squares, in Settings 1, 3, and 5. The proposed method has the lowest error with respect to both norms in the two sparse settings (1 and 3), while its non-sparse version and partial least squares are better in the non-sparse setting (5). Also shown in Figure~\ref{fig:simulbeta} are the estimated curves for each of the four methods for Settings 1, 3, and 5, along with the median in solid black and the true discriminant curve in a dot-dashed black line. The median curve of the estimators from the proposed method resembles the true function the most in Settings 1 and 3, successfully identifying the non-signal regions. The estimators with a ridge penalty show larger variability than other methods, which suggests that a ridge-type regularization in functional classification does not necessarily reduce variance, unlike in ridge regression.

\begin{figure}[tbph]
\caption{For each real data example, 20 sample curves from each class are shown along with average curves. Solid and dotted curves represent curves from different classes.}\label{fig:realdata}
       \begin{subfigure}[b]{0.33\textwidth}
        \centering
        \includegraphics[width = \textwidth, height = 1.3in]{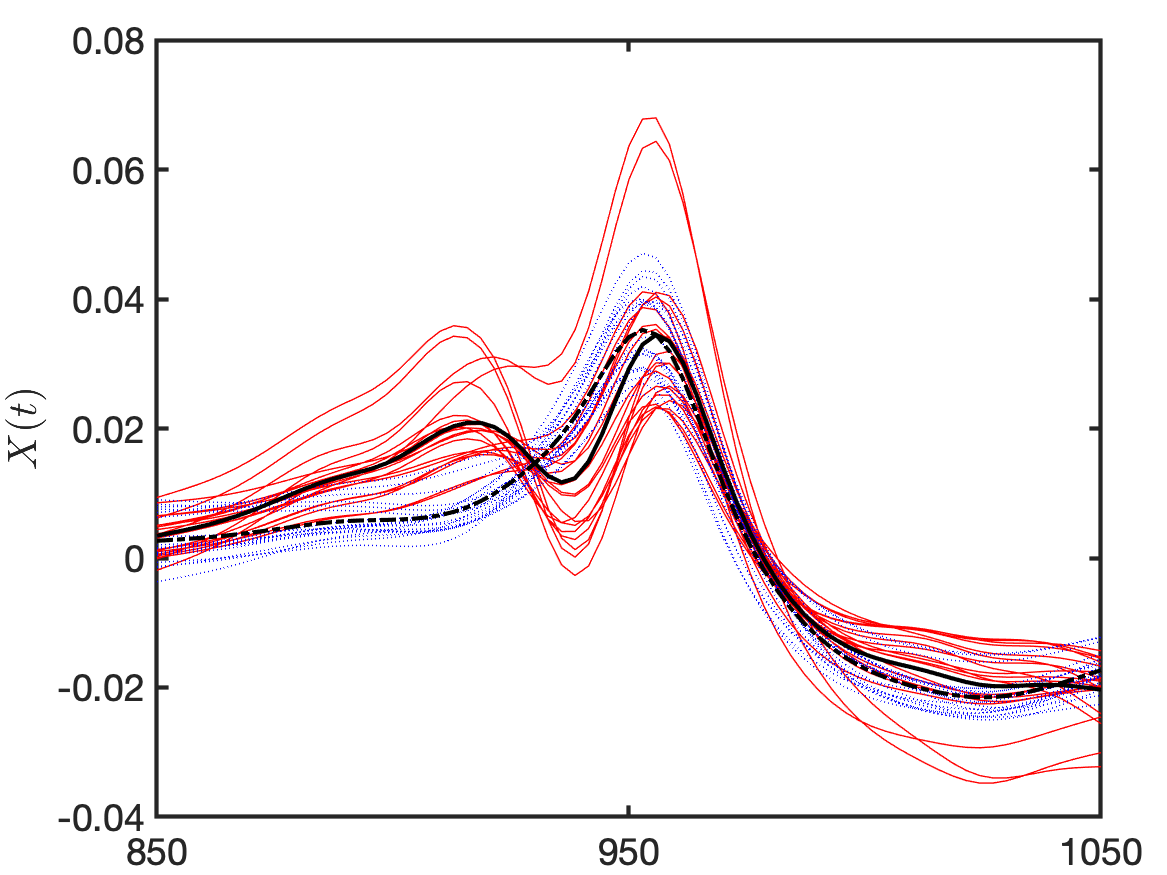}
        \caption{Tecator}
    \end{subfigure}%
         \begin{subfigure}[b]{0.33\textwidth}
        \centering
        \includegraphics[width = \textwidth, height = 1.3in]{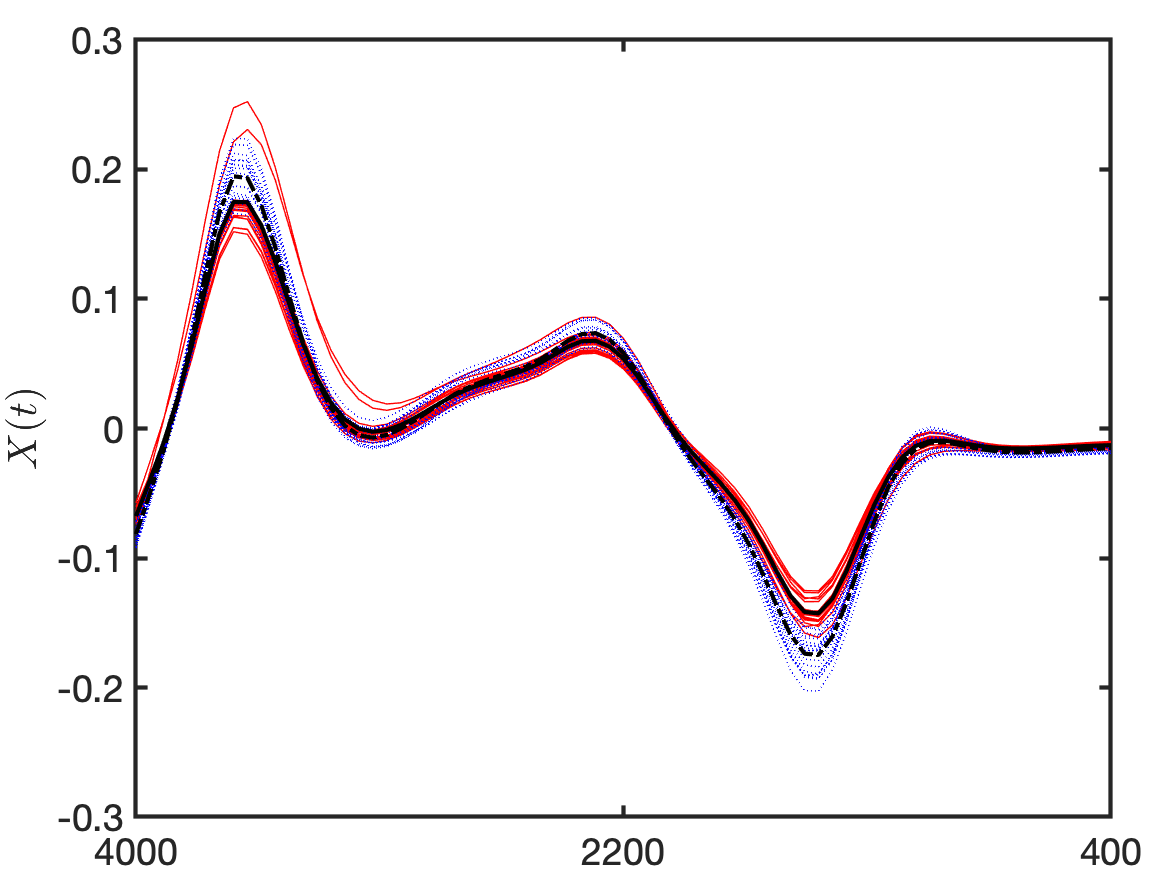}
        \caption{Wine}
            \end{subfigure}%
    \begin{subfigure}[b]{0.33\textwidth}
        \centering
        \includegraphics[width = \textwidth, height = 1.3in]{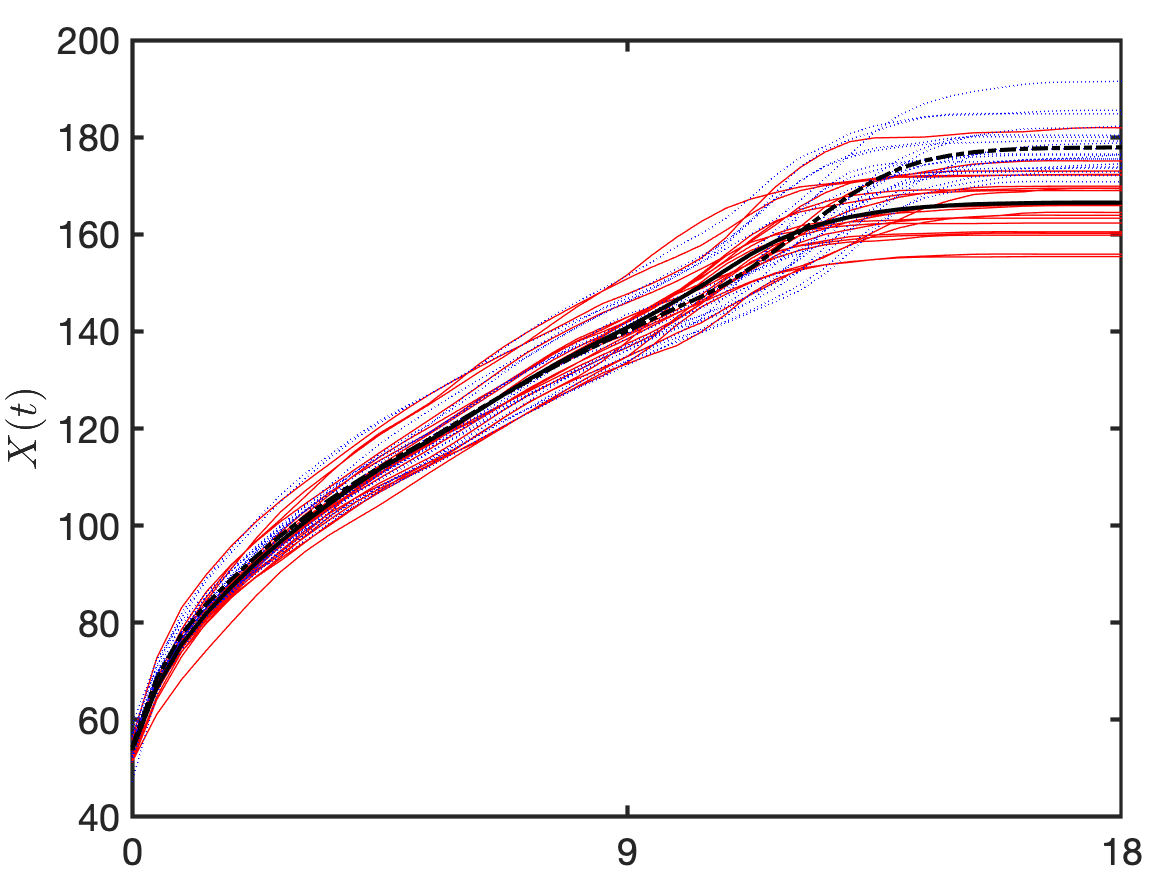}
        \caption{Growth}
    \end{subfigure}\\   
    \begin{subfigure}[b]{0.33\textwidth}
        \centering
        \includegraphics[width = \textwidth, height = 1.3in]{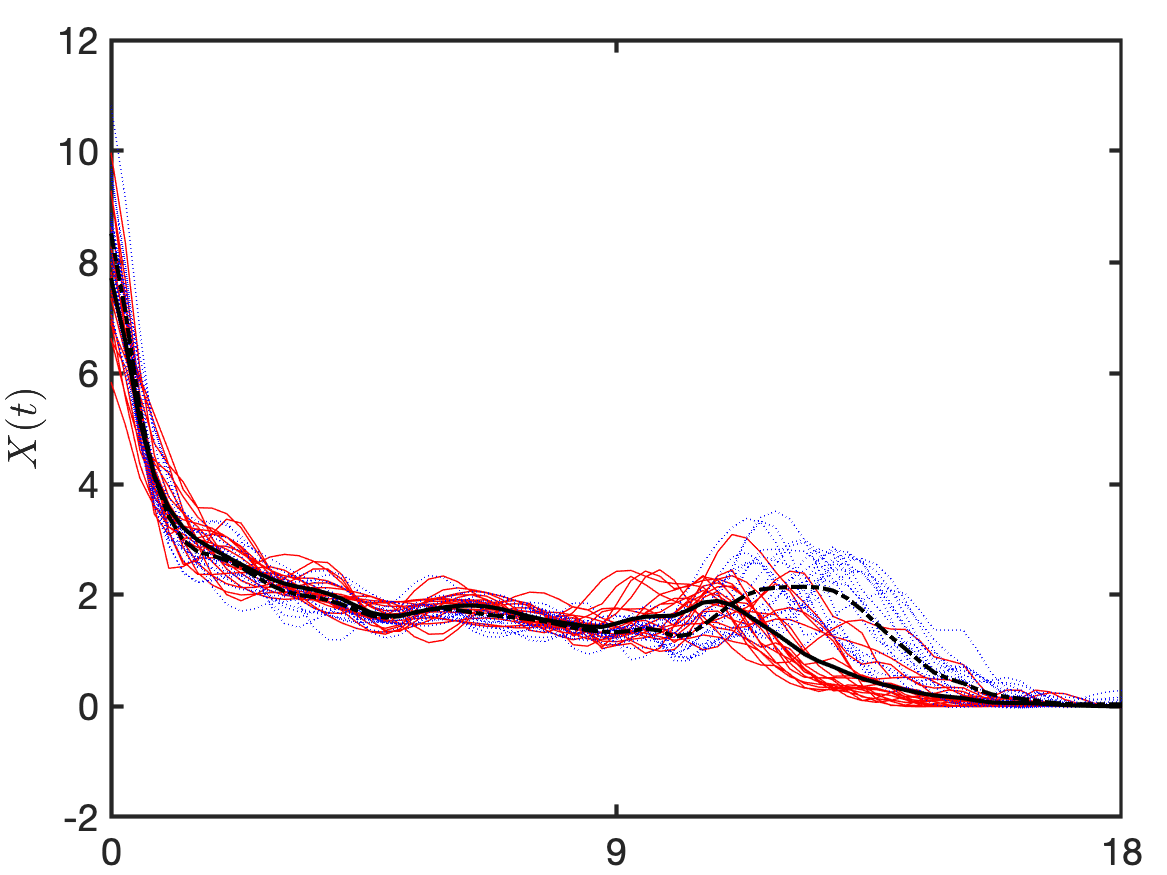}
        \caption{dGrowth}
    \end{subfigure}%
    \begin{subfigure}[b]{0.33\textwidth}
        \centering
        \includegraphics[width = \textwidth, height = 1.3in]{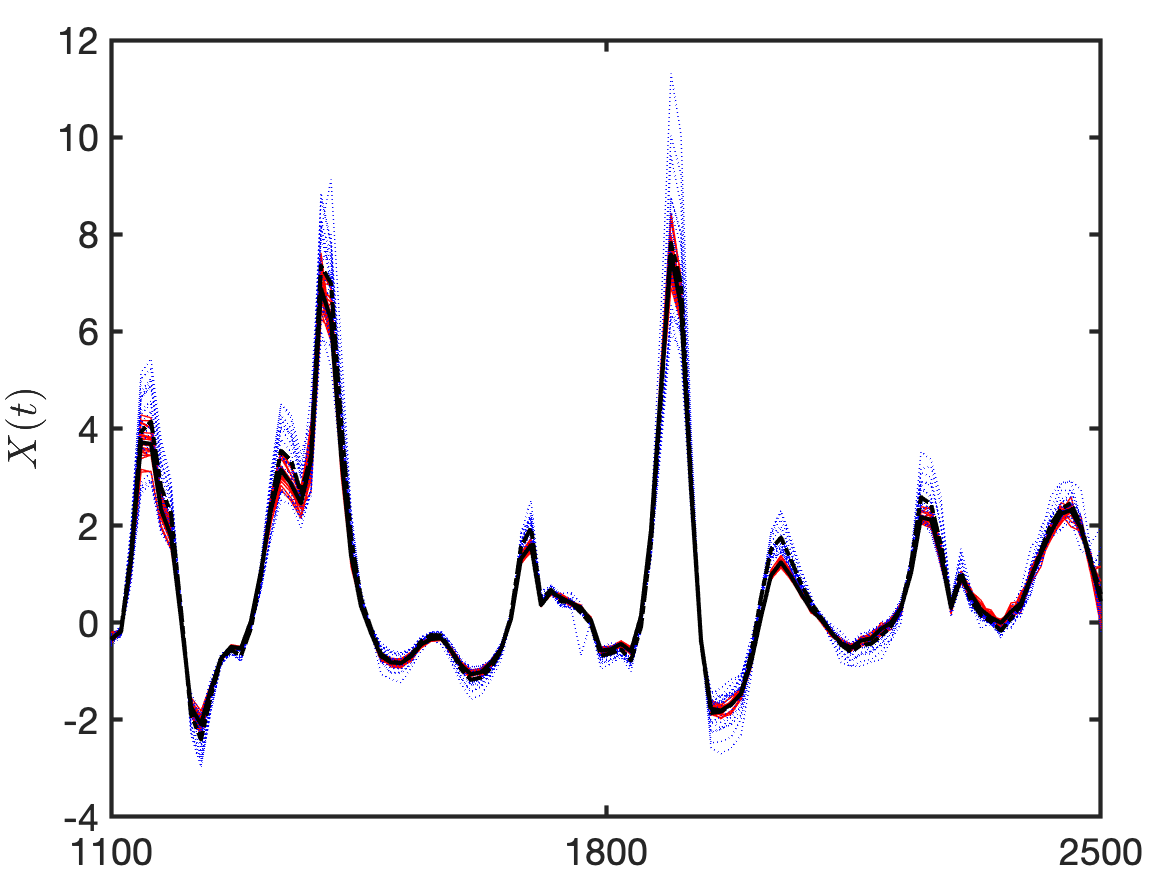}
        \caption{Wheat}
    \end{subfigure}%
    \begin{subfigure}[b]{0.33\textwidth}
        \centering
        \includegraphics[width = \textwidth, height = 1.3in]{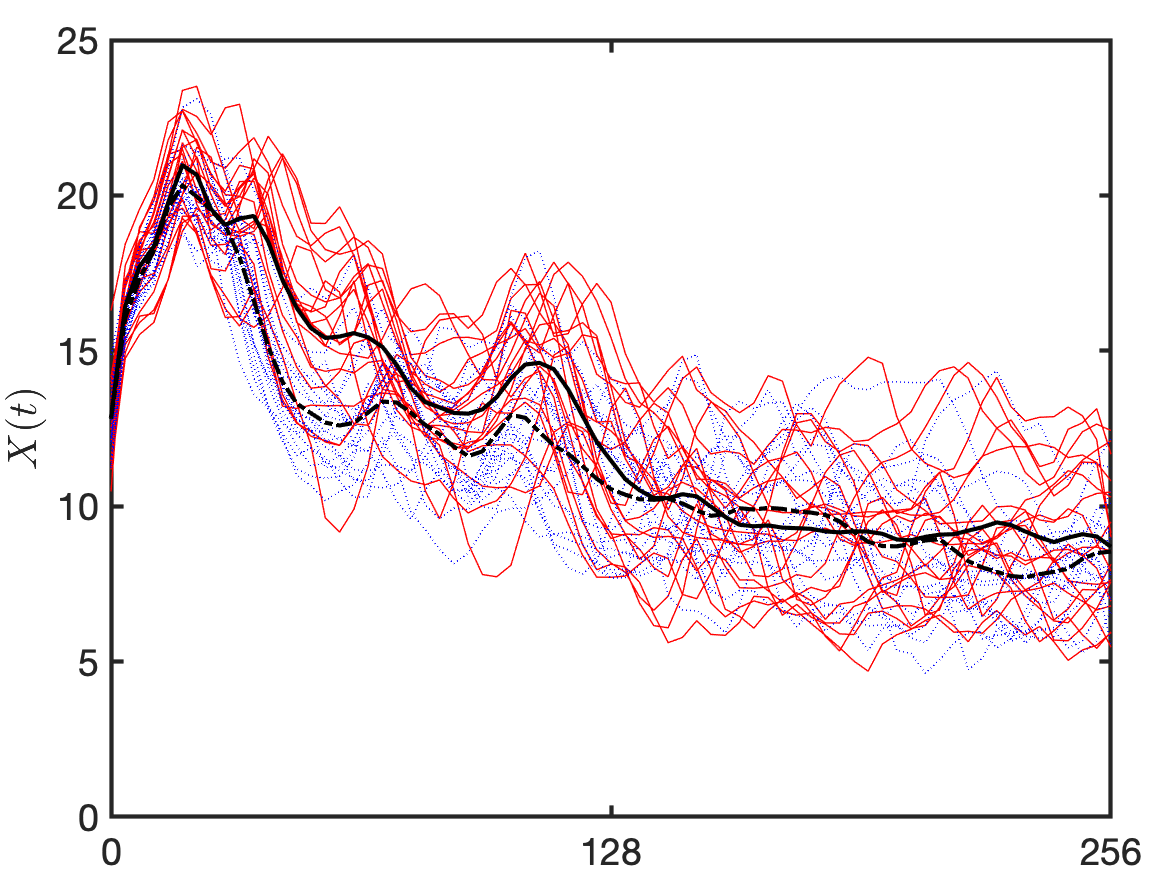}
        \caption{Phoneme}
    \end{subfigure}%
\end{figure}

\subsection{Real data examples}

\begin{figure}[tbph]
  \caption{Estimated discriminant curves for real data examples from 30 repetitions. The median curve is shown as a thicker black solid line.}\label{fig:realbeta}
  \footnotesize{
       \begin{subfigure}[b]{0.245\textwidth}
        \centering
        \includegraphics[width = \textwidth, height = 1in]{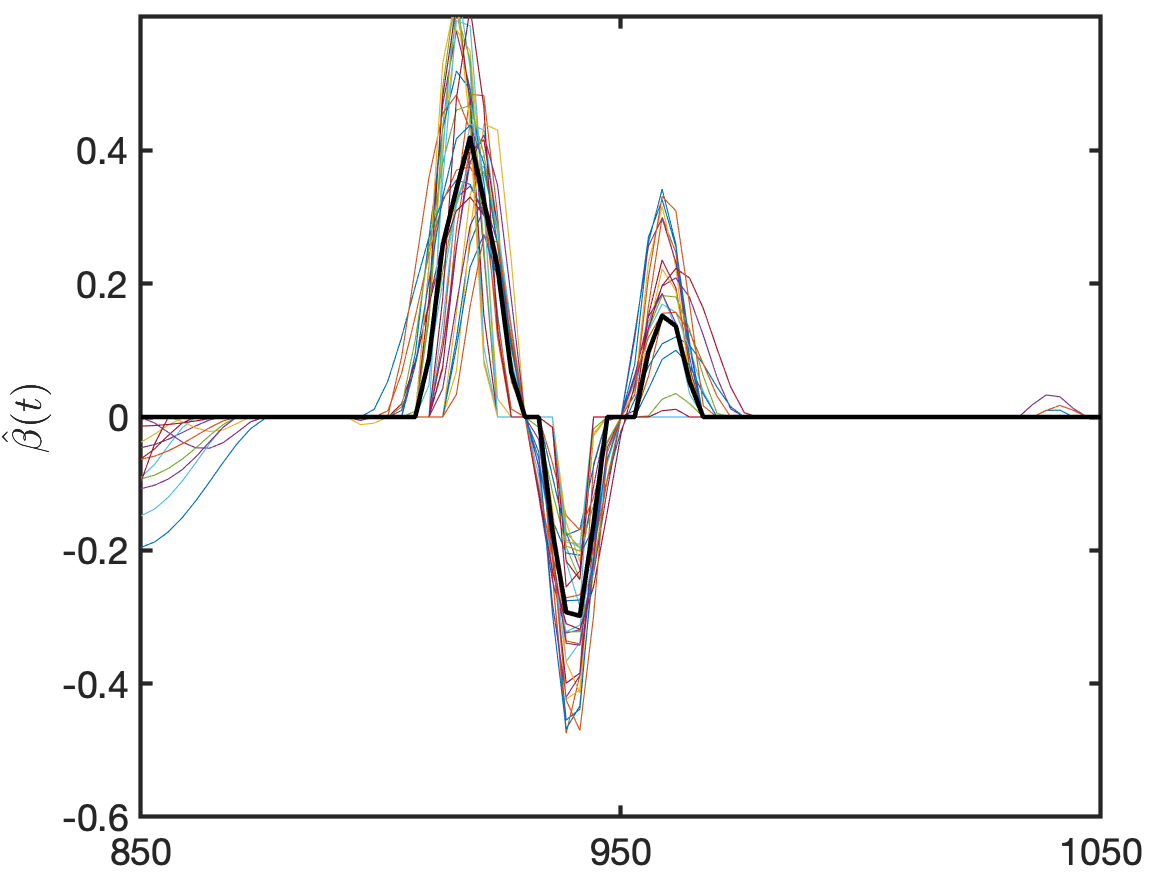}
        \caption*{Tecator - SFLDA}
        \end{subfigure}
         \begin{subfigure}[b]{0.245\textwidth}
        \centering
        \includegraphics[width = \textwidth, height = 1in]{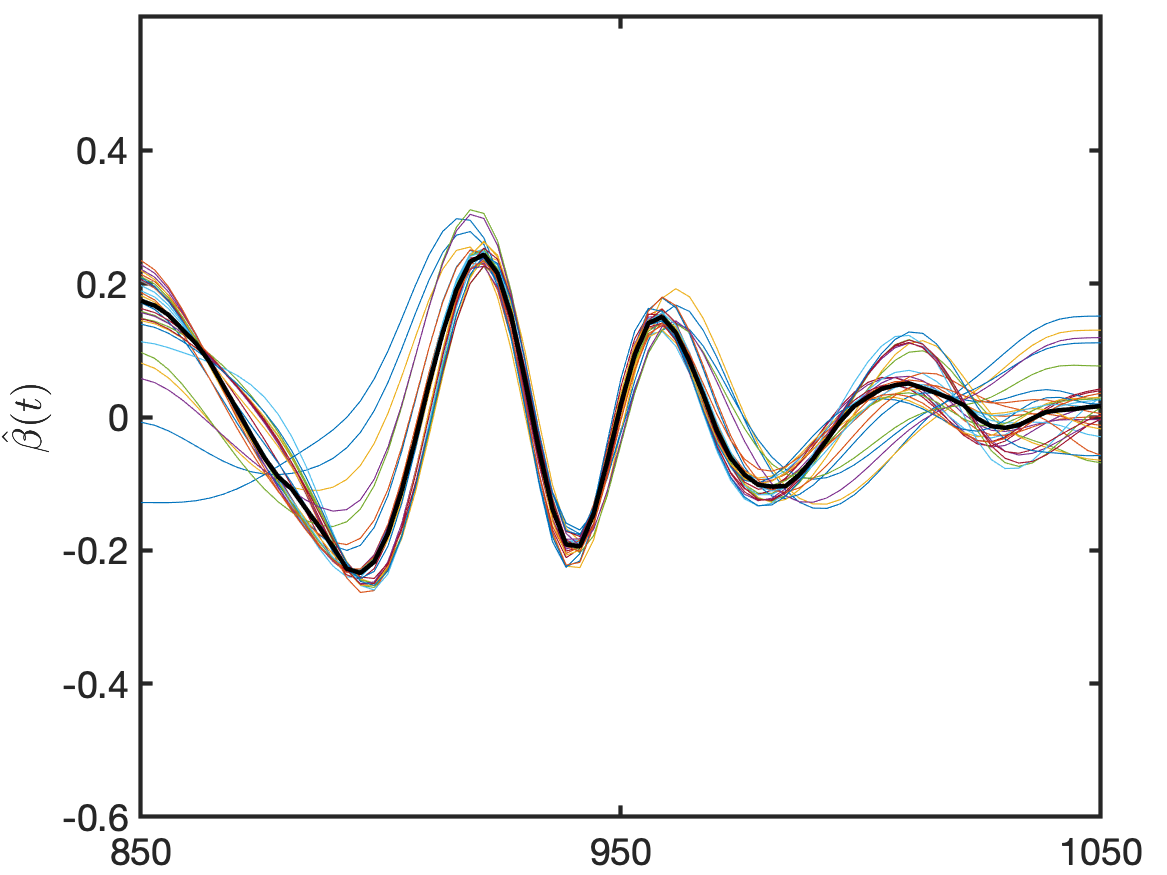}
        \caption*{FLDA}
        \end{subfigure}
               \begin{subfigure}[b]{0.245\textwidth}
        \centering
        \includegraphics[width = \textwidth, height = 1in]{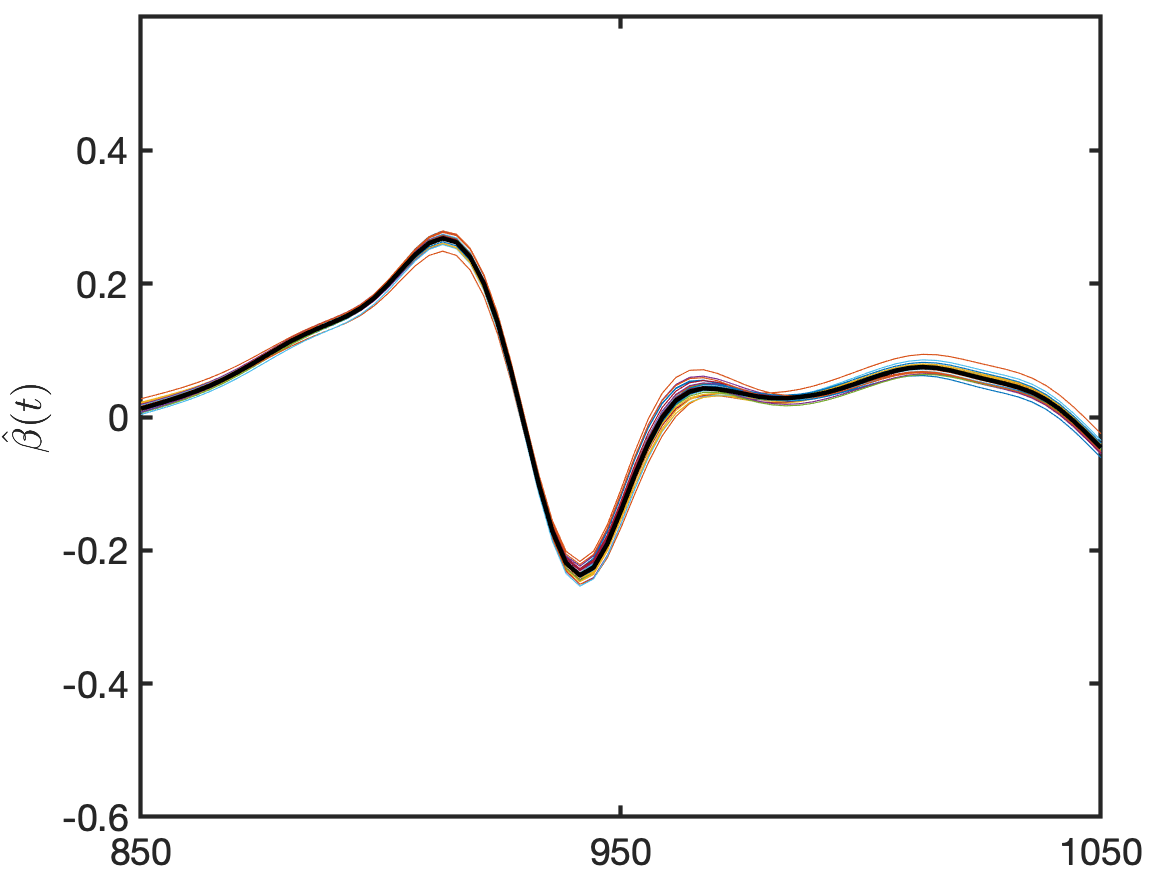}
        \caption*{RFLDA}
        \end{subfigure}
        \begin{subfigure}[b]{0.245\textwidth}
        \centering
        \includegraphics[width = \textwidth, height = 1in]{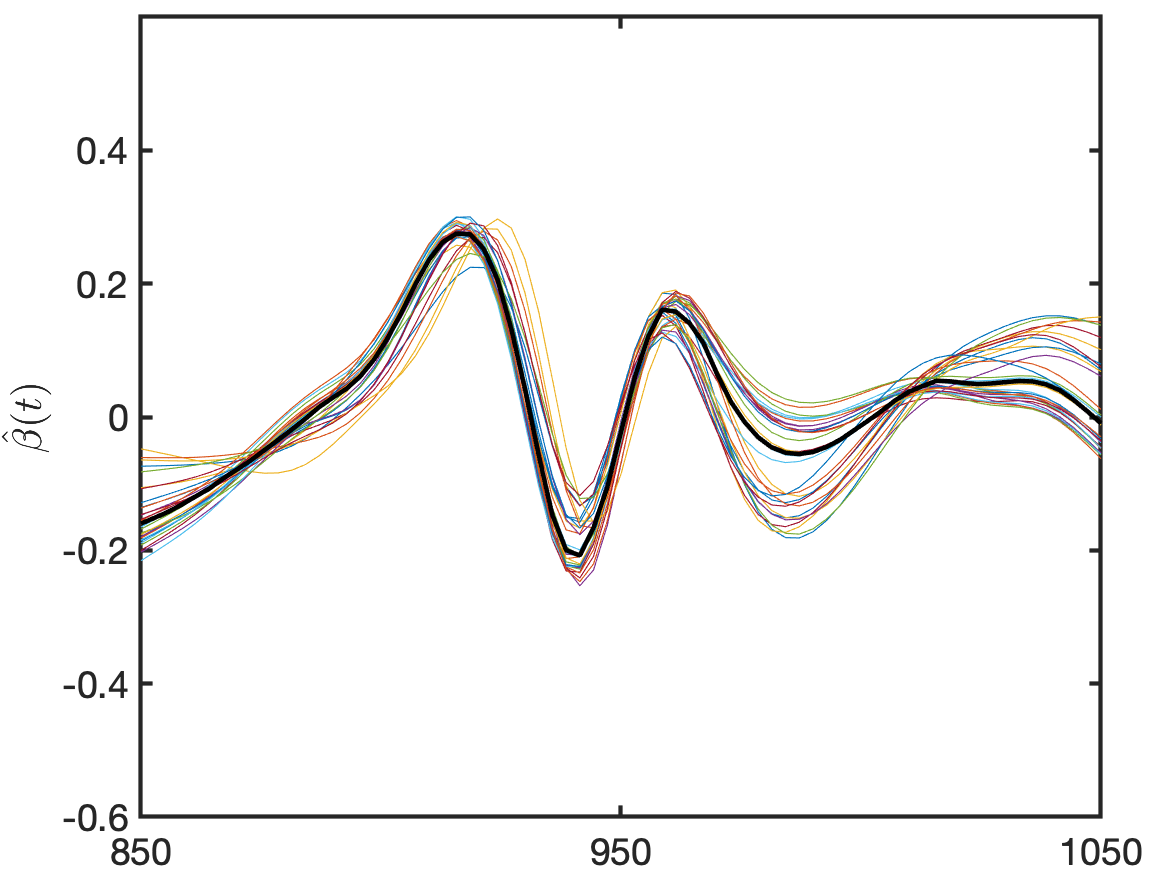}
        \caption*{PLS}
        \end{subfigure}\\
       \begin{subfigure}[b]{0.245\textwidth}
        \centering
        \includegraphics[width = \textwidth, height = 1in]{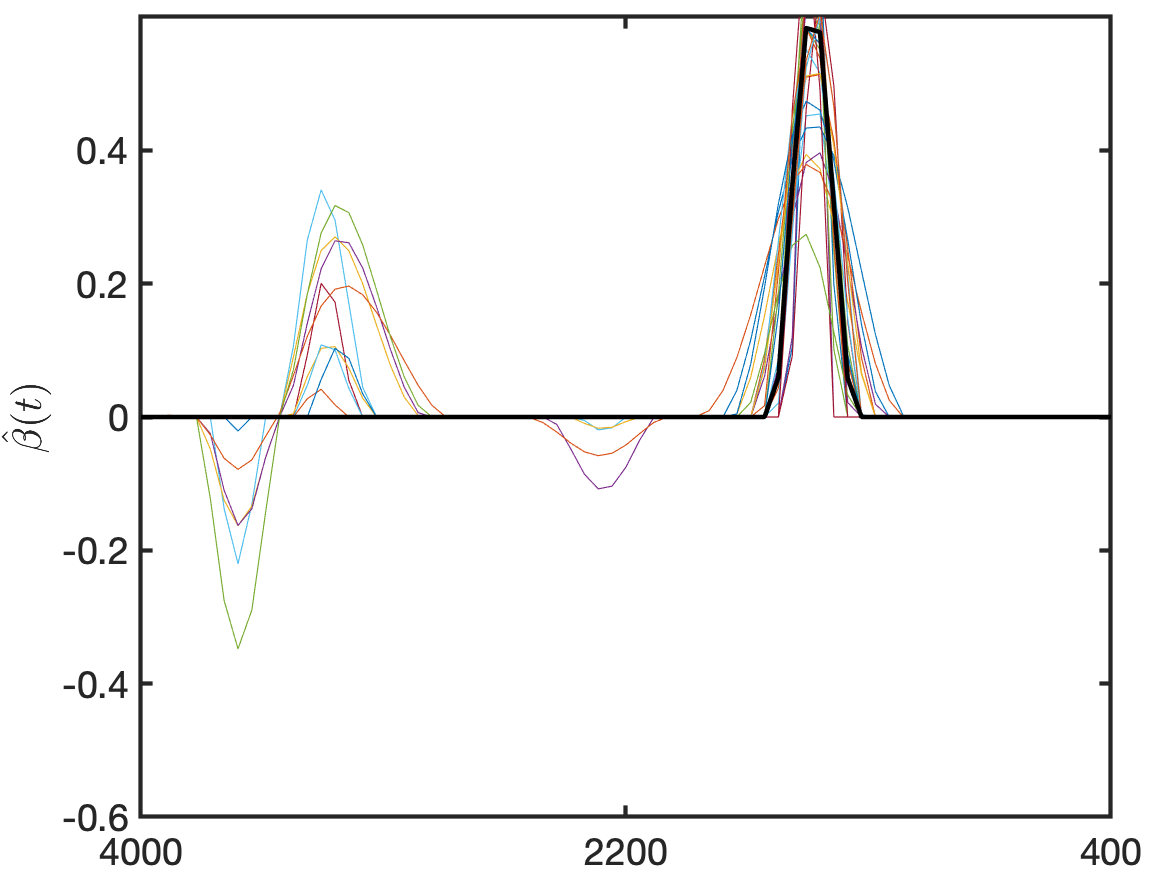}
        \caption*{Wine - SFLDA}
        \end{subfigure}
       \begin{subfigure}[b]{0.245\textwidth}
        \centering
        \includegraphics[width = \textwidth, height = 1in]{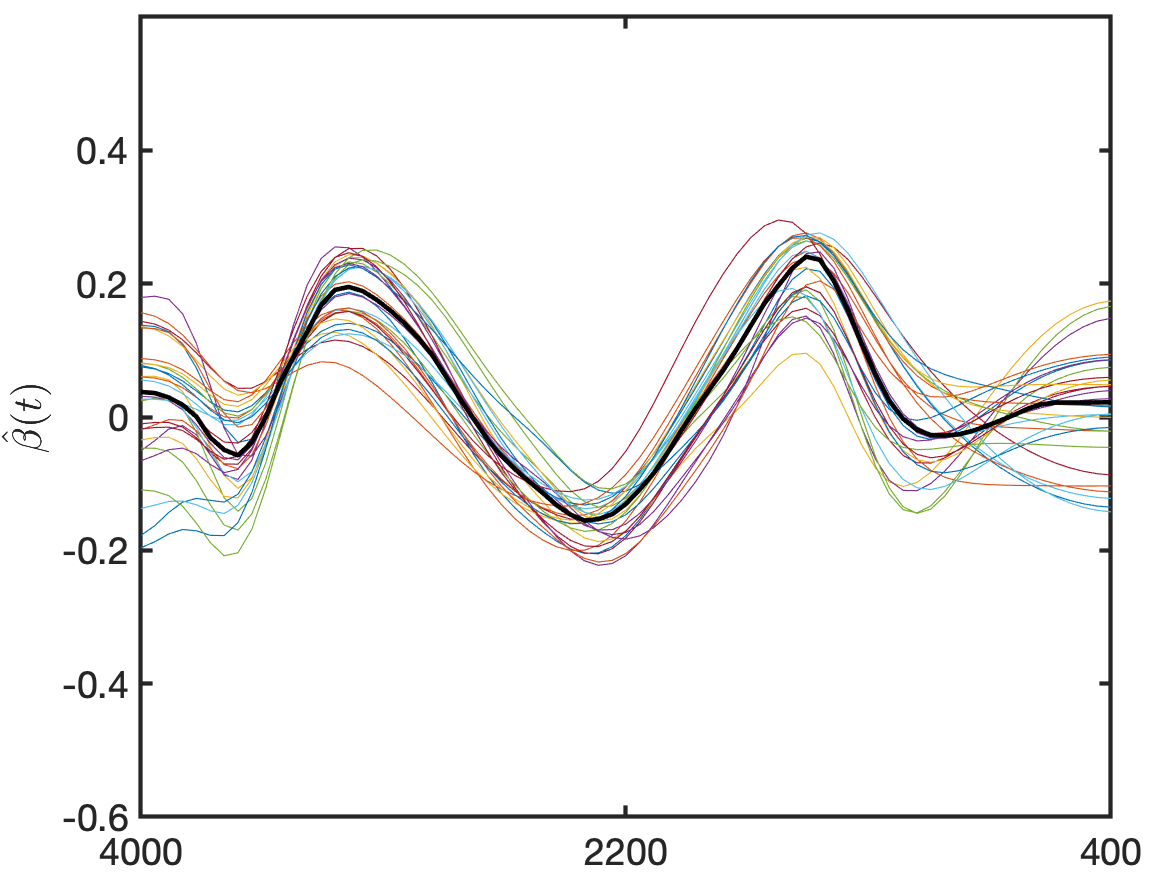}
        \caption*{FLDA}
        \end{subfigure}
   \begin{subfigure}[b]{0.245\textwidth}
        \centering
        \includegraphics[width = \textwidth, height = 1in]{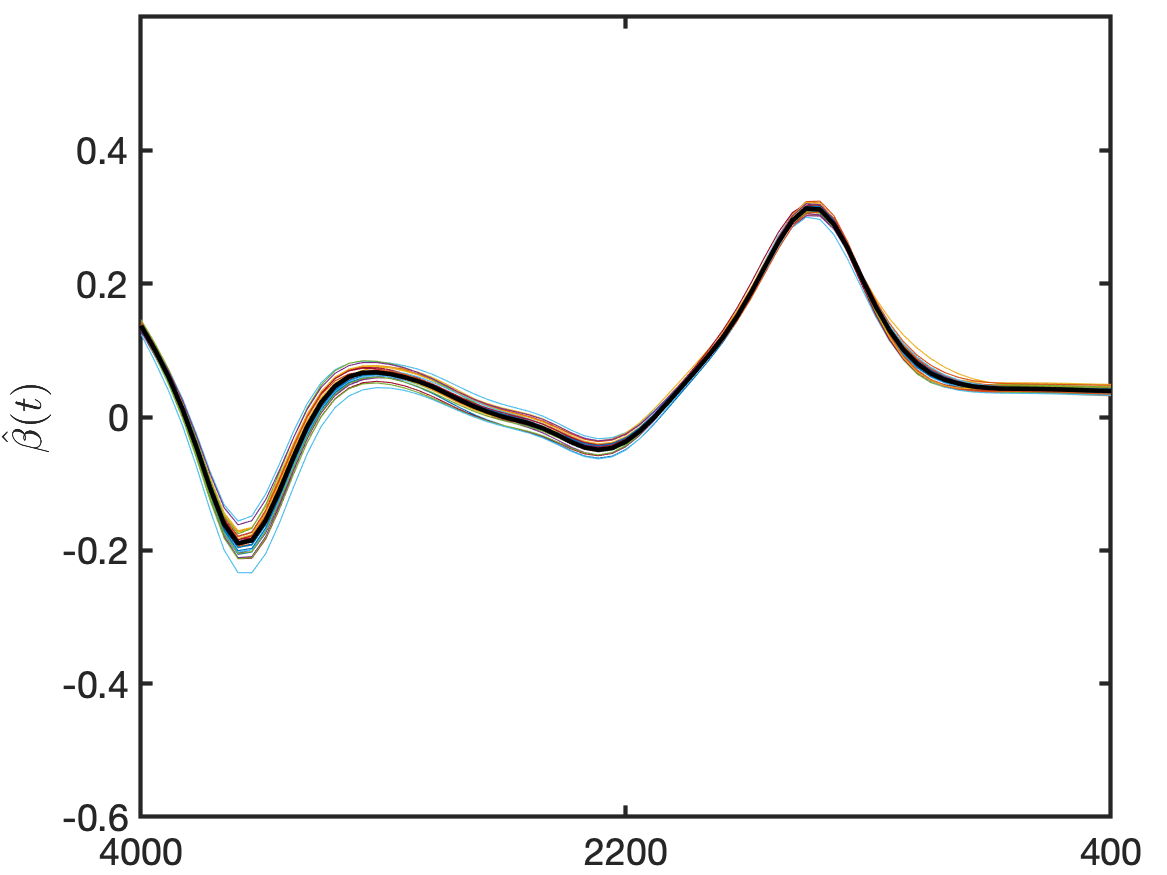}
        \caption*{RFLDA}
        \end{subfigure}
   \begin{subfigure}[b]{0.245\textwidth}
        \centering
        \includegraphics[width = \textwidth, height = 1in]{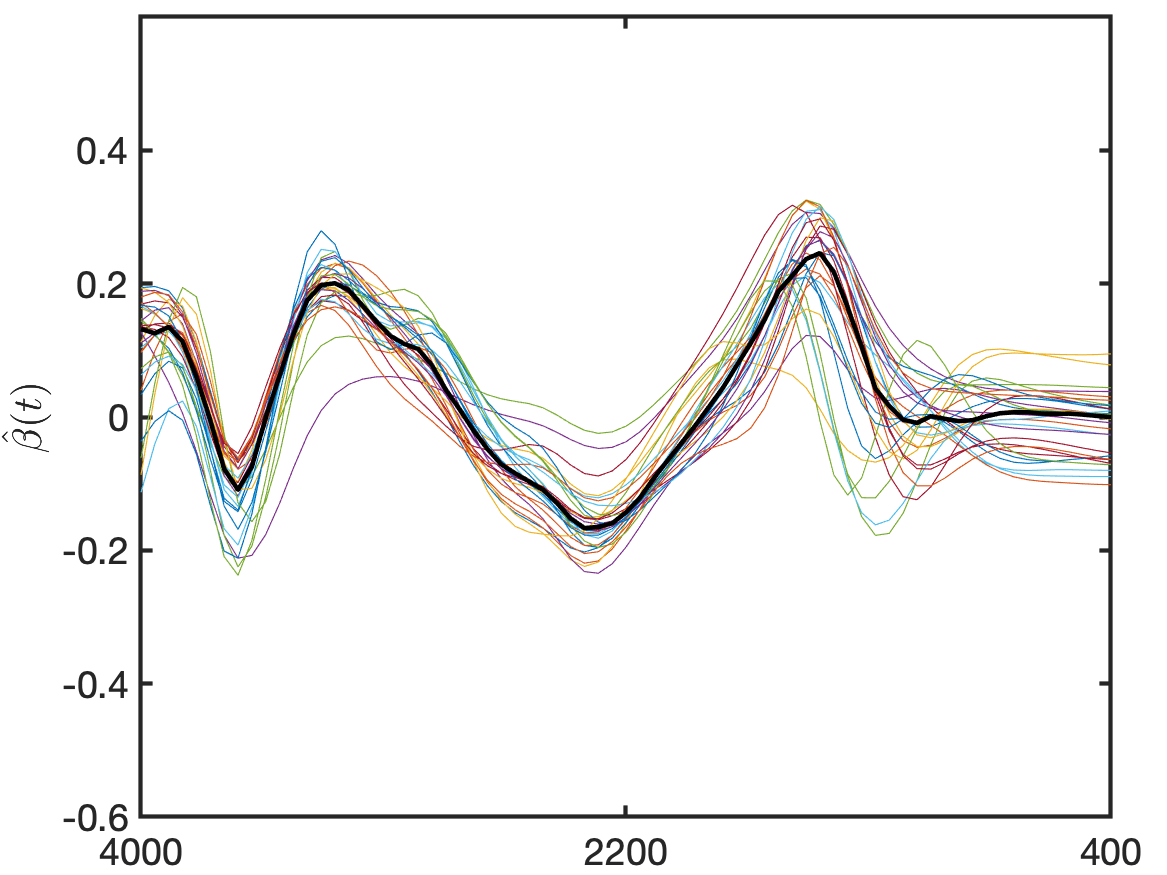}
        \caption*{PLS}
        \end{subfigure}\\
    \begin{subfigure}[b]{0.245\textwidth}
        \centering
        \includegraphics[width = \textwidth, height = 1in]{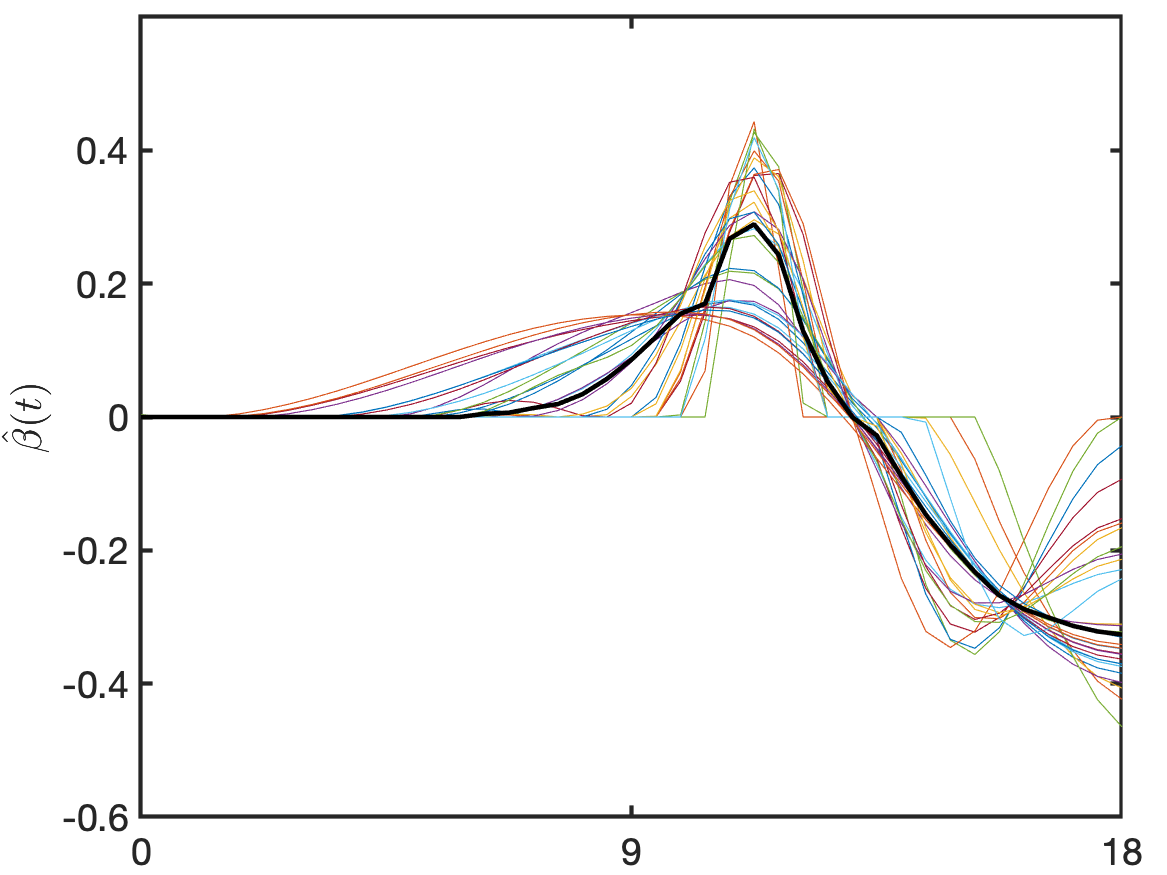}
        \caption*{Growth - SFLDA}
        \end{subfigure}
     \begin{subfigure}[b]{0.245\textwidth}
        \centering
        \includegraphics[width = \textwidth, height = 1in]{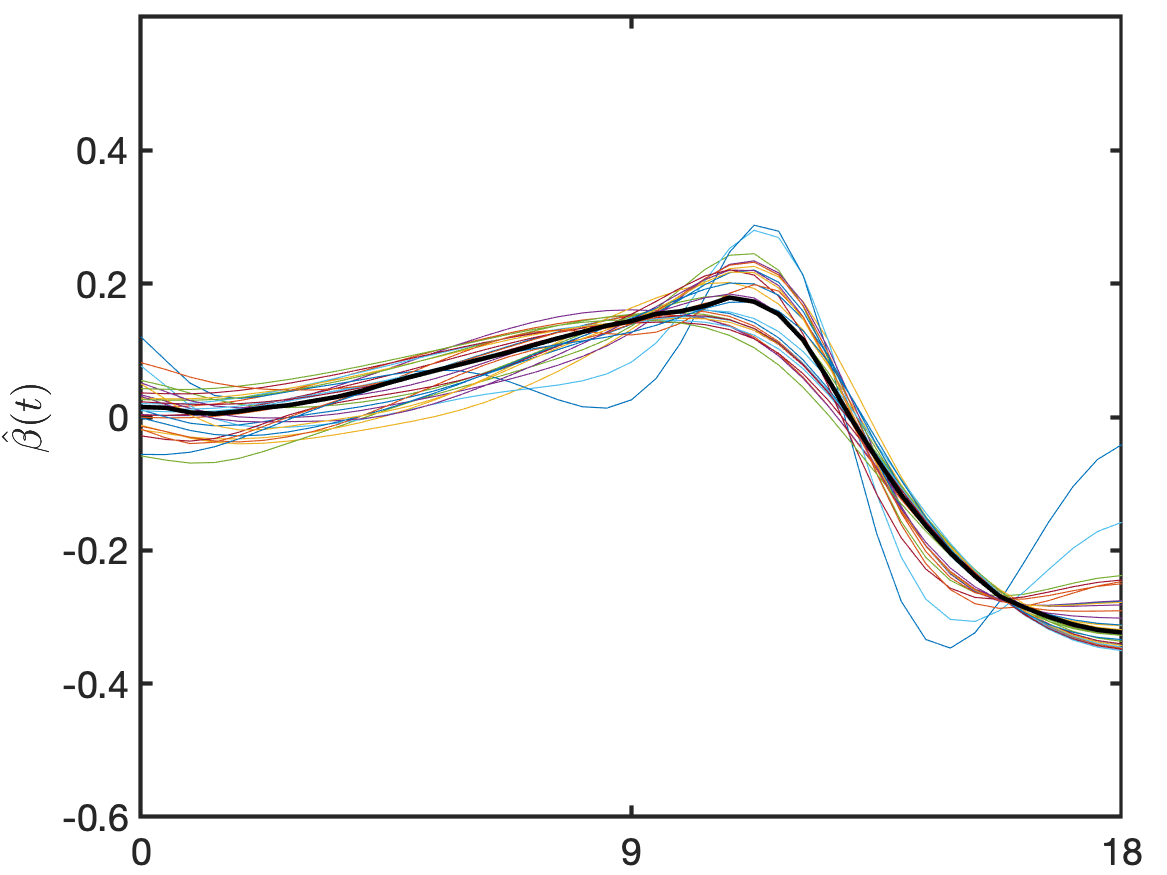}
        \caption*{FLDA}
        \end{subfigure}
    \begin{subfigure}[b]{0.245\textwidth}
        \centering
        \includegraphics[width = \textwidth, height = 1in]{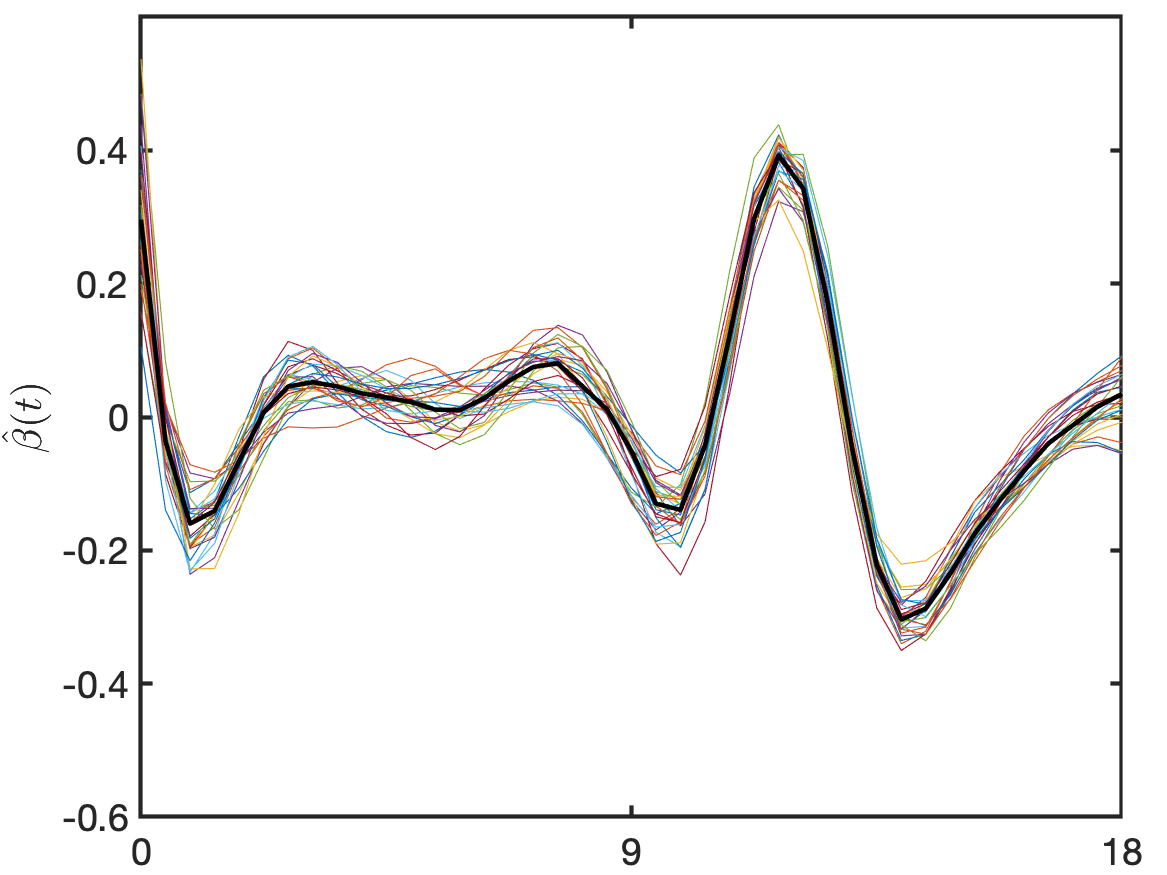}
        \caption*{RFLDA}
        \end{subfigure}
    \begin{subfigure}[b]{0.245\textwidth}
        \centering
        \includegraphics[width = \textwidth, height = 1in]{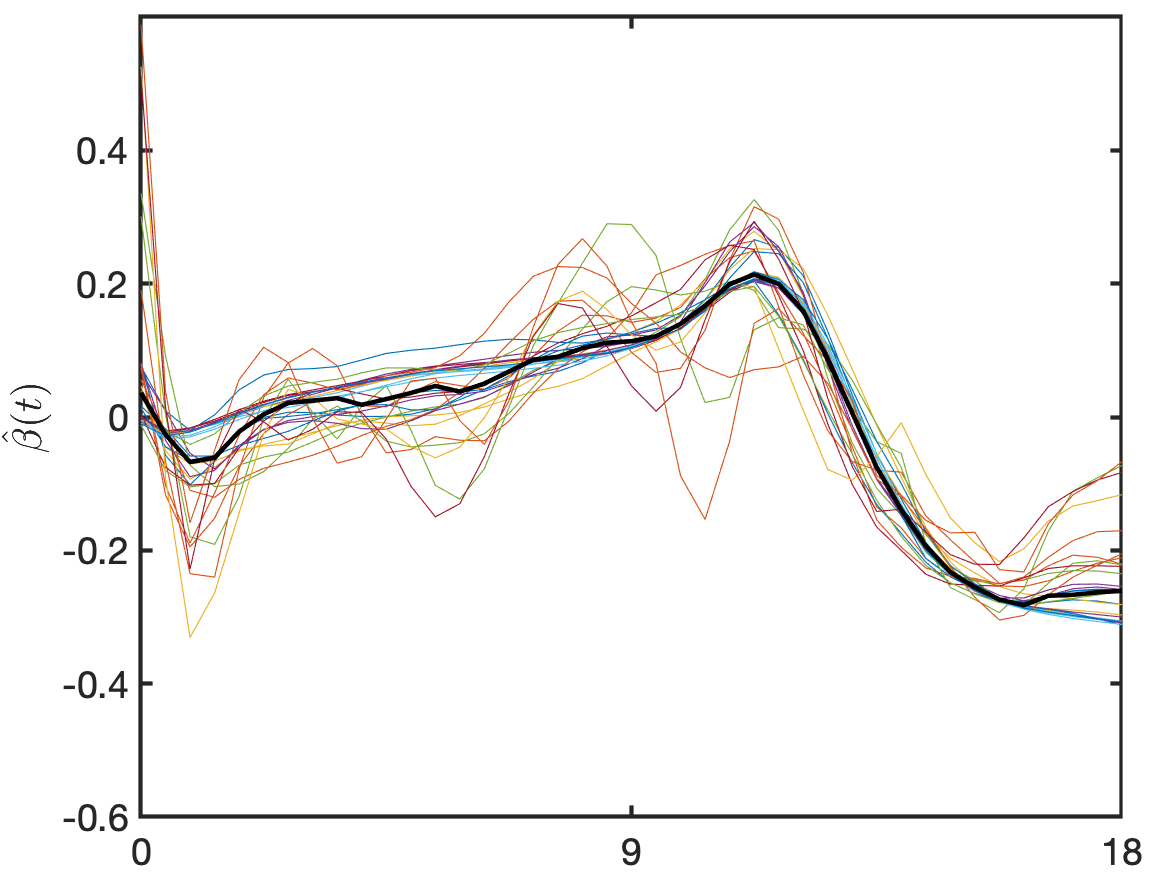}
        \caption*{PLS}
        \end{subfigure}\\
    \begin{subfigure}[b]{0.245\textwidth}
        \centering
        \includegraphics[width = \textwidth, height = 1in]{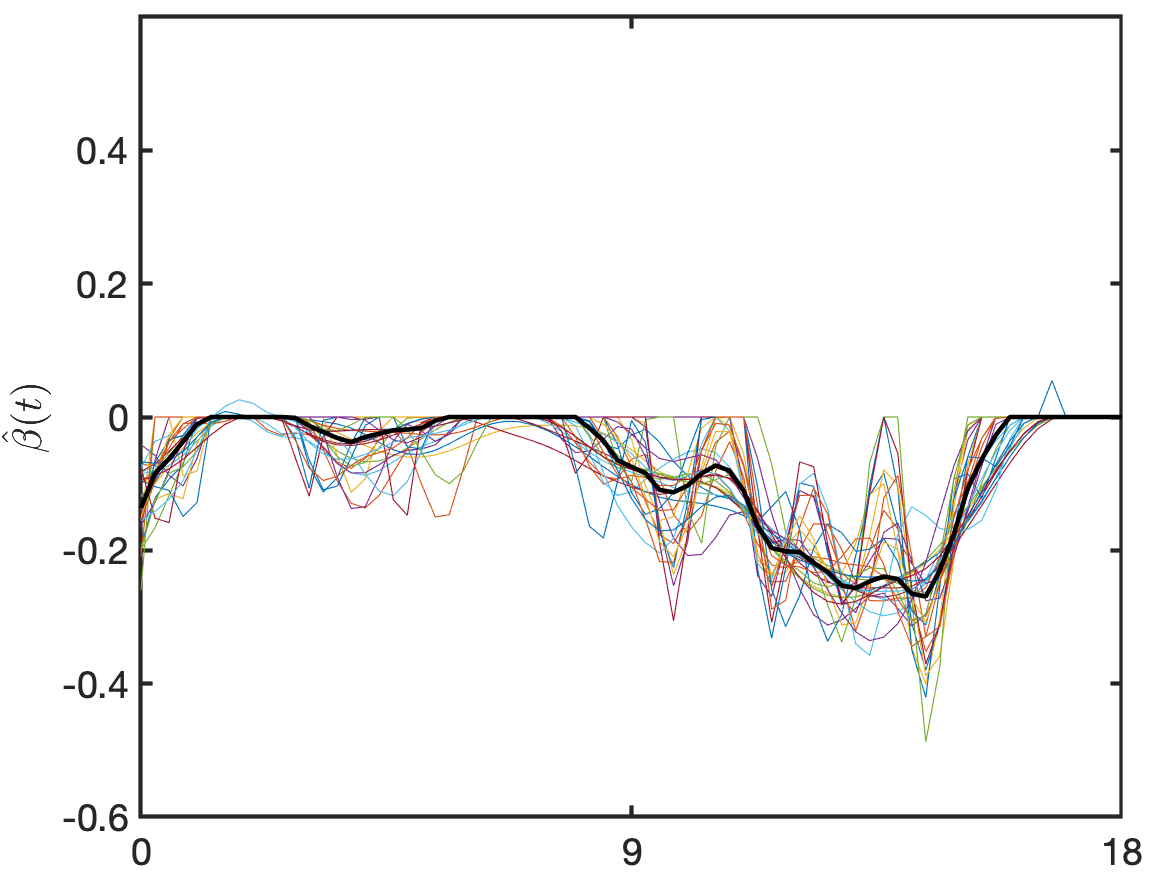}
        \caption*{Dgrowth - SFLDA}
        \end{subfigure}
    \begin{subfigure}[b]{0.245\textwidth}
        \centering
        \includegraphics[width = \textwidth, height = 1in]{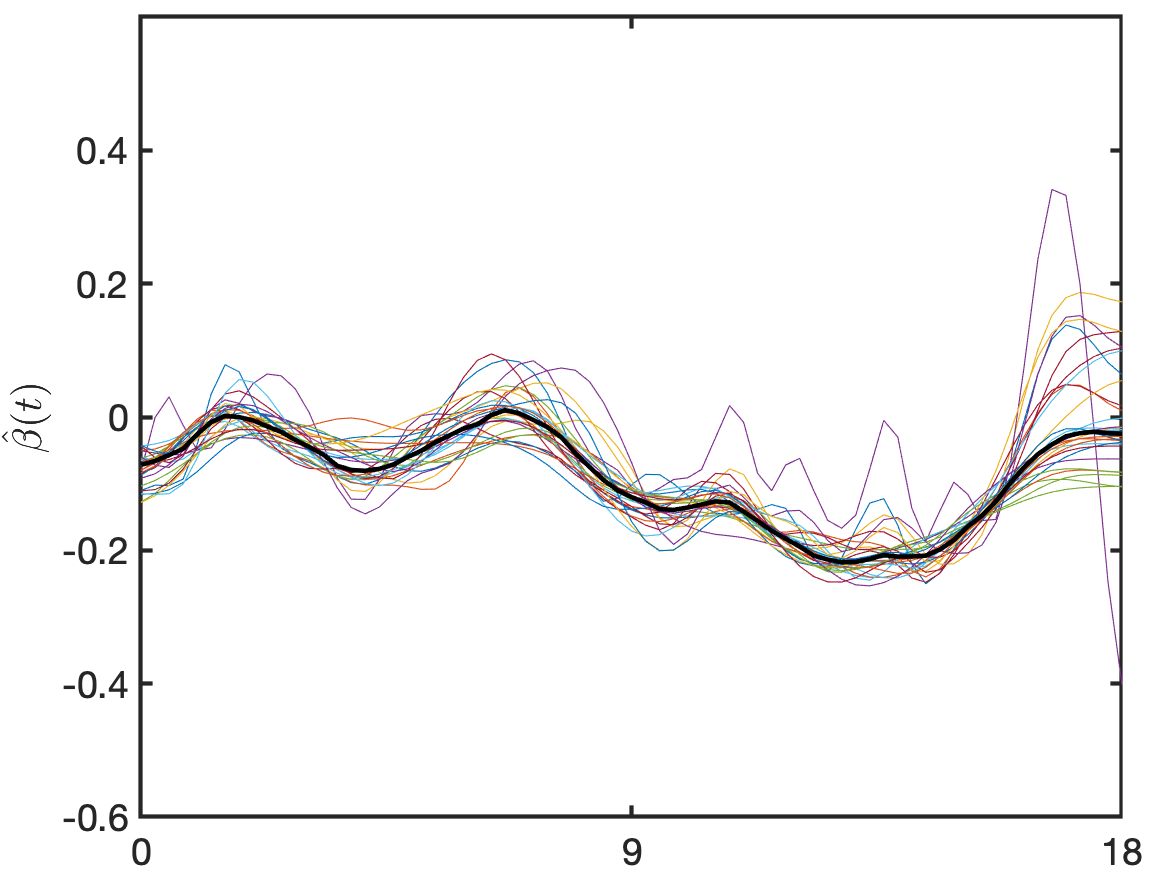}
        \caption*{FLDA}
        \end{subfigure}
            \begin{subfigure}[b]{0.245\textwidth}
        \centering
        \includegraphics[width = \textwidth, height = 1in]{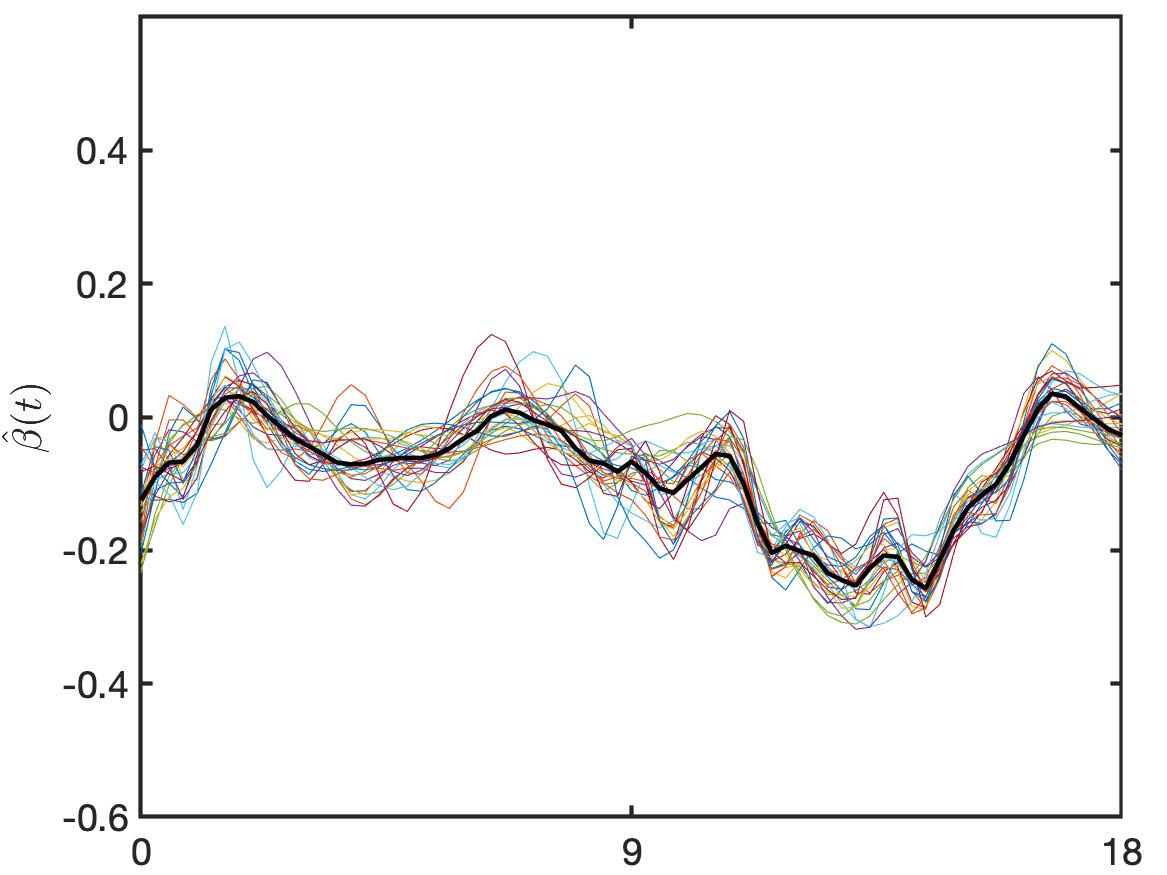}
        \caption*{RFLDA}
        \end{subfigure}
            \begin{subfigure}[b]{0.245\textwidth}
        \centering
        \includegraphics[width = \textwidth, height = 1in]{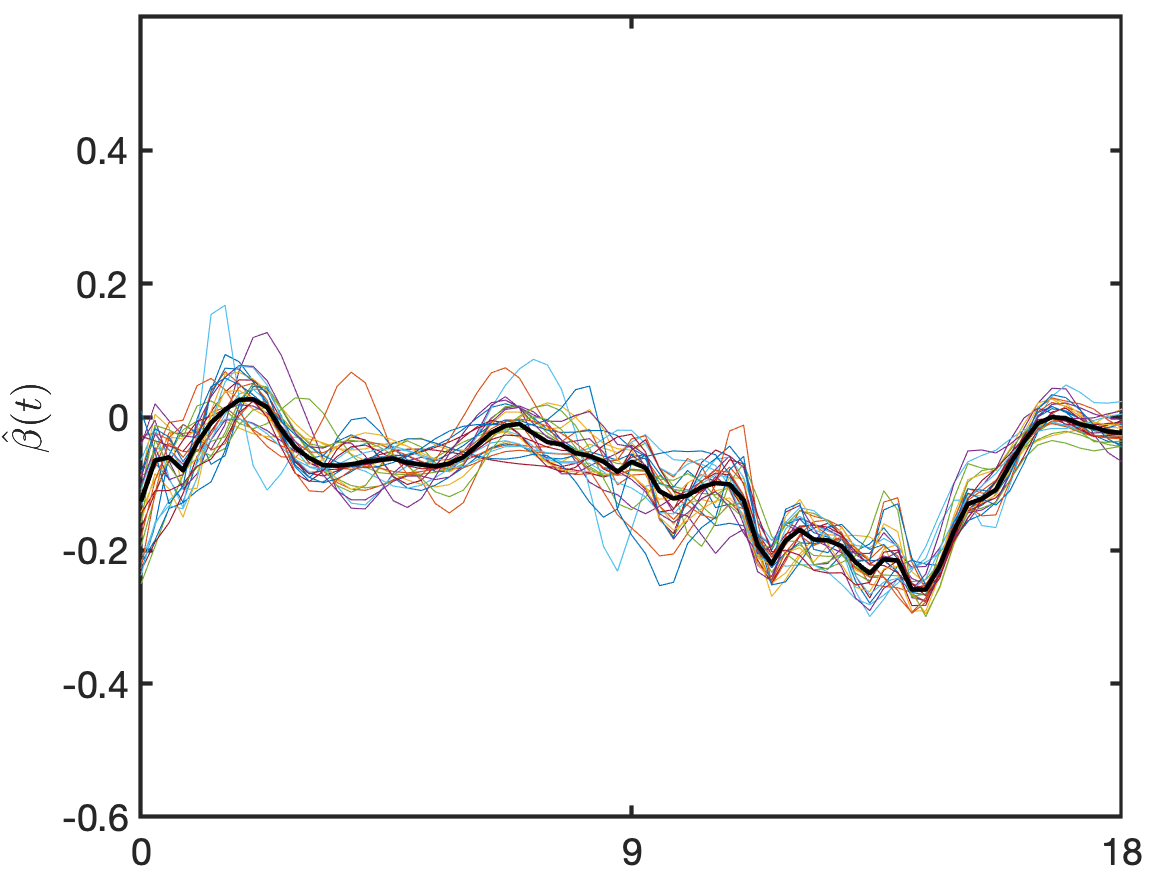}
        \caption*{PLS}
        \end{subfigure}\\
      \begin{subfigure}[b]{0.245\textwidth}
        \centering
        \includegraphics[width = \textwidth, height = 1in]{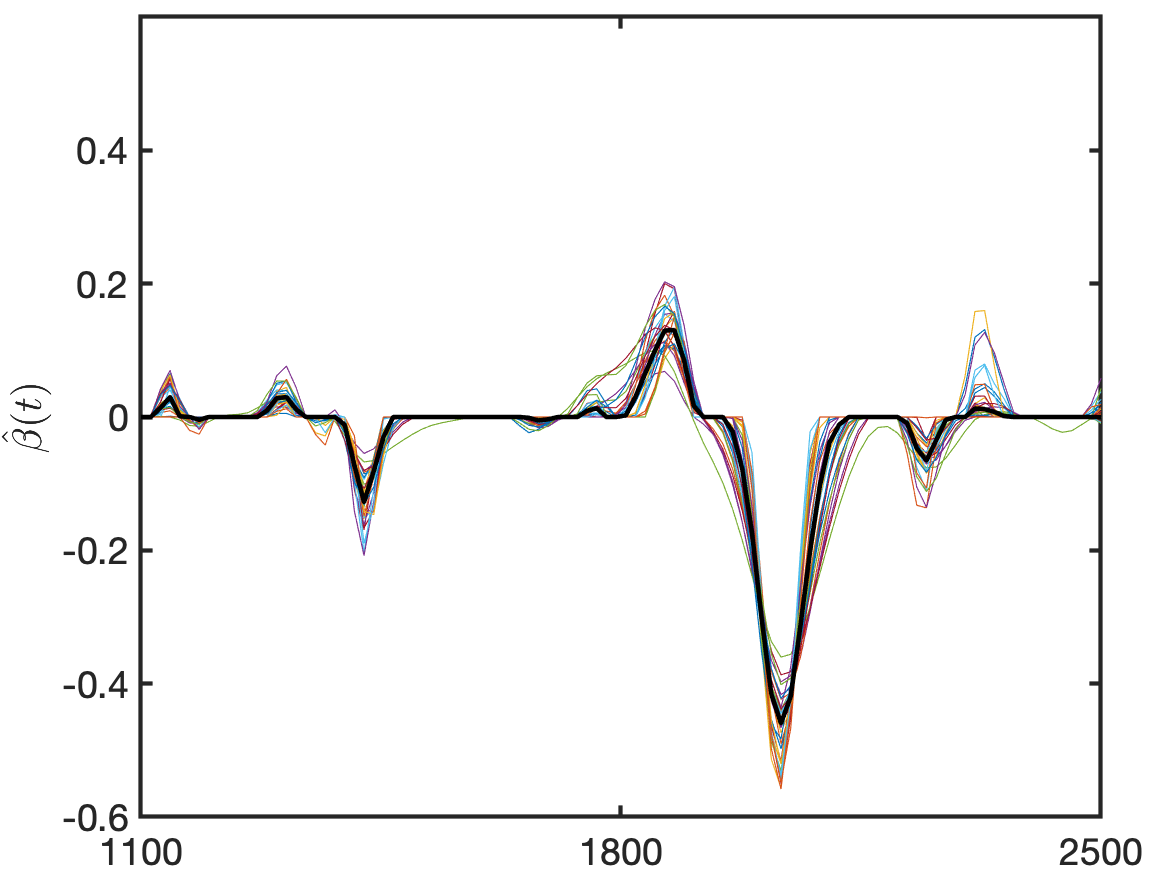}
        \caption*{Wheat - SFLDA}
        \end{subfigure}
       \begin{subfigure}[b]{0.245\textwidth}
        \centering
        \includegraphics[width = \textwidth, height = 1in]{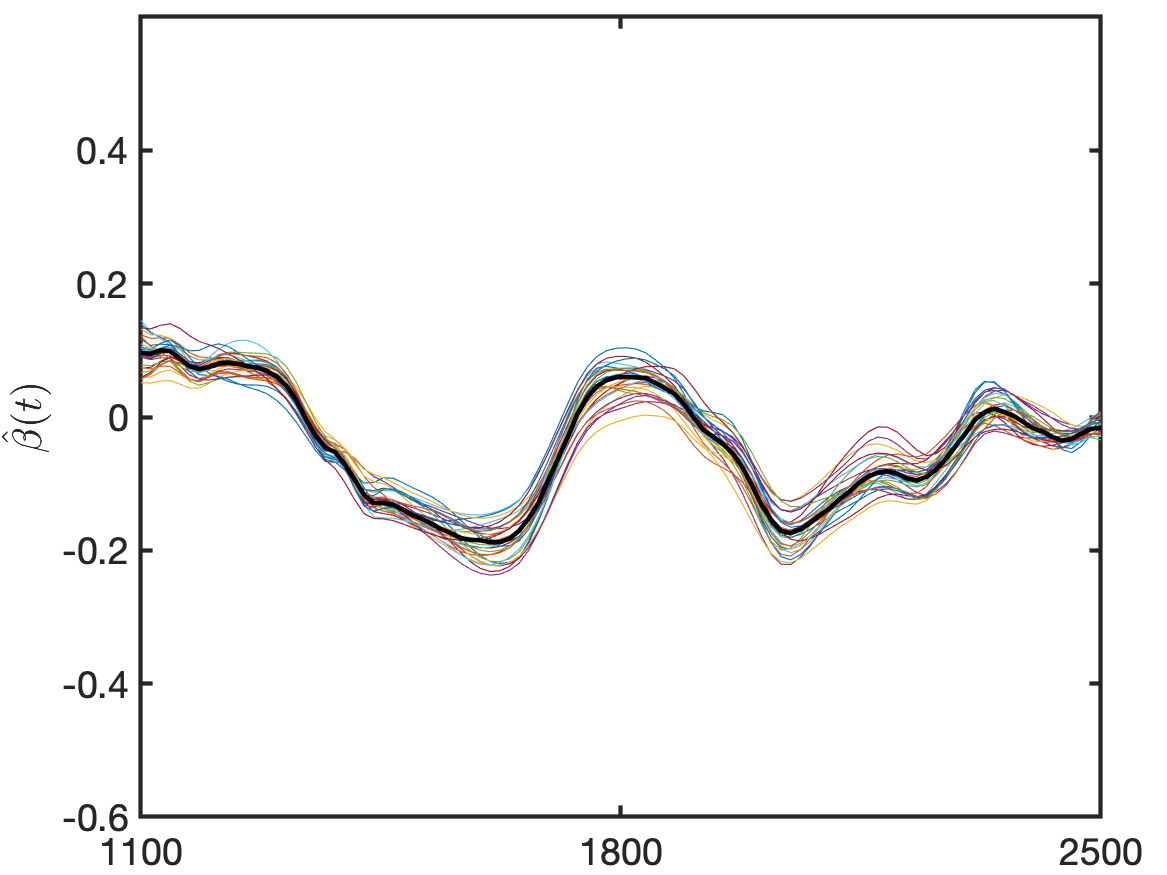}
        \caption*{FLDA}
        \end{subfigure}            
        \begin{subfigure}[b]{0.245\textwidth}
        \centering
        \includegraphics[width = \textwidth, height = 1in]{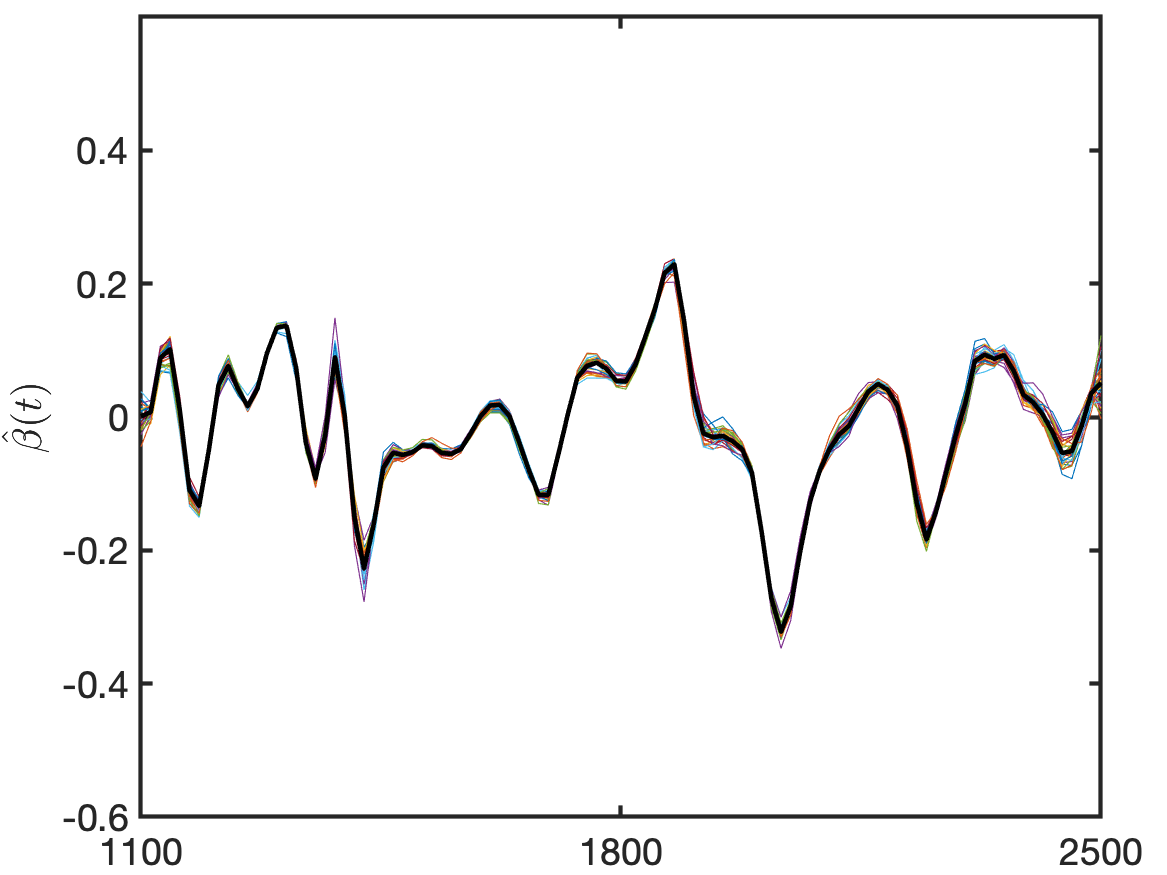}
        \caption*{RFLDA}
        \end{subfigure}            
        \begin{subfigure}[b]{0.245\textwidth}
        \centering
        \includegraphics[width = \textwidth, height = 1in]{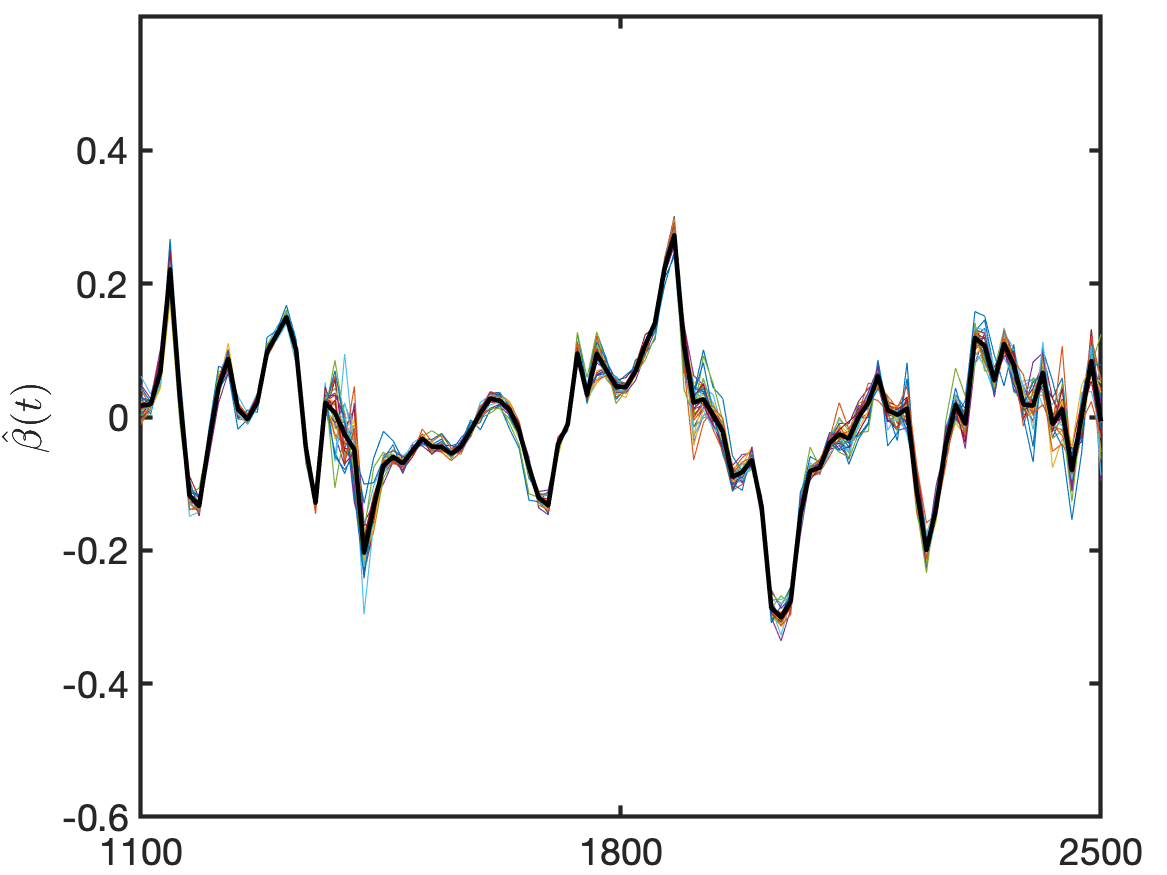}
        \caption*{PLS}
        \end{subfigure}\\
				 \begin{subfigure}[b]{0.245\textwidth}
        \centering
        \includegraphics[width = \textwidth, height = 1in]{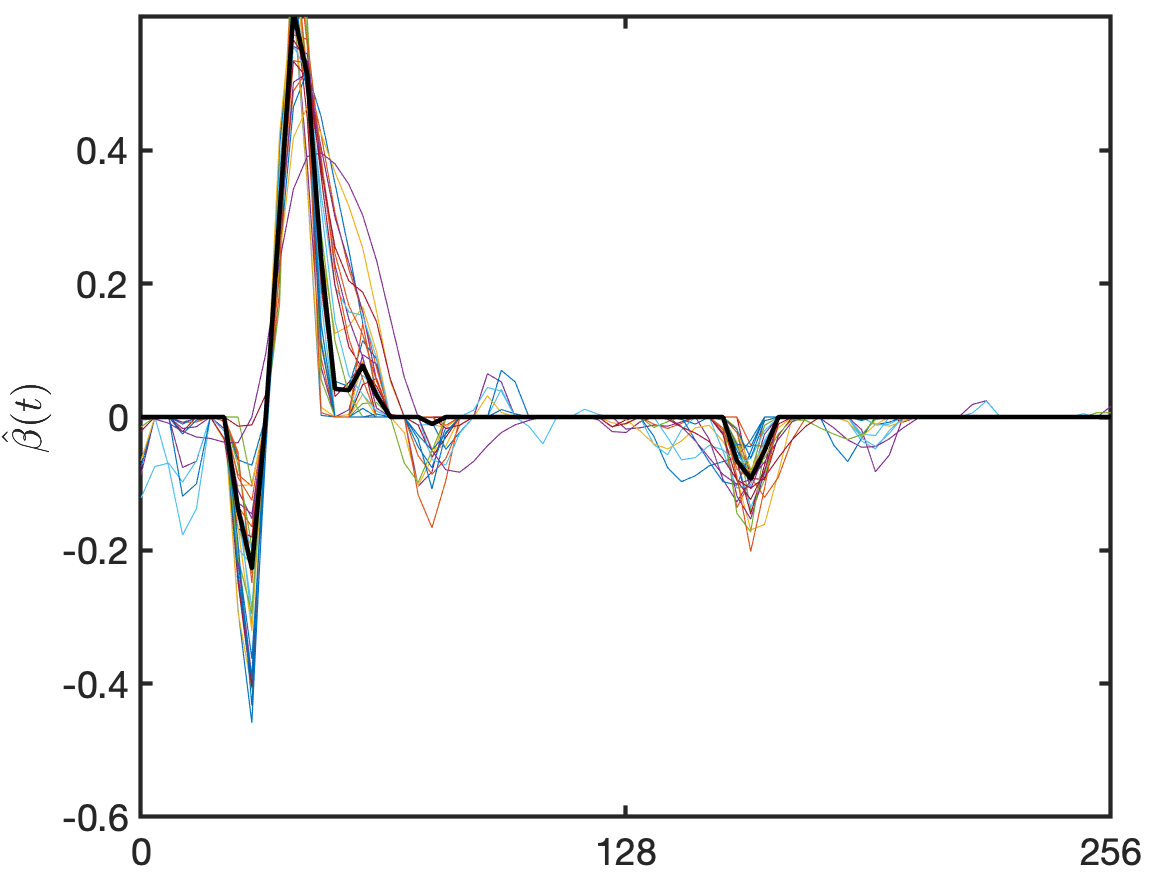}
			\caption*{Phoneme - SFLDA}
        \end{subfigure}
       \begin{subfigure}[b]{0.245\textwidth}
        \centering
        \includegraphics[width = \textwidth, height = 1in]{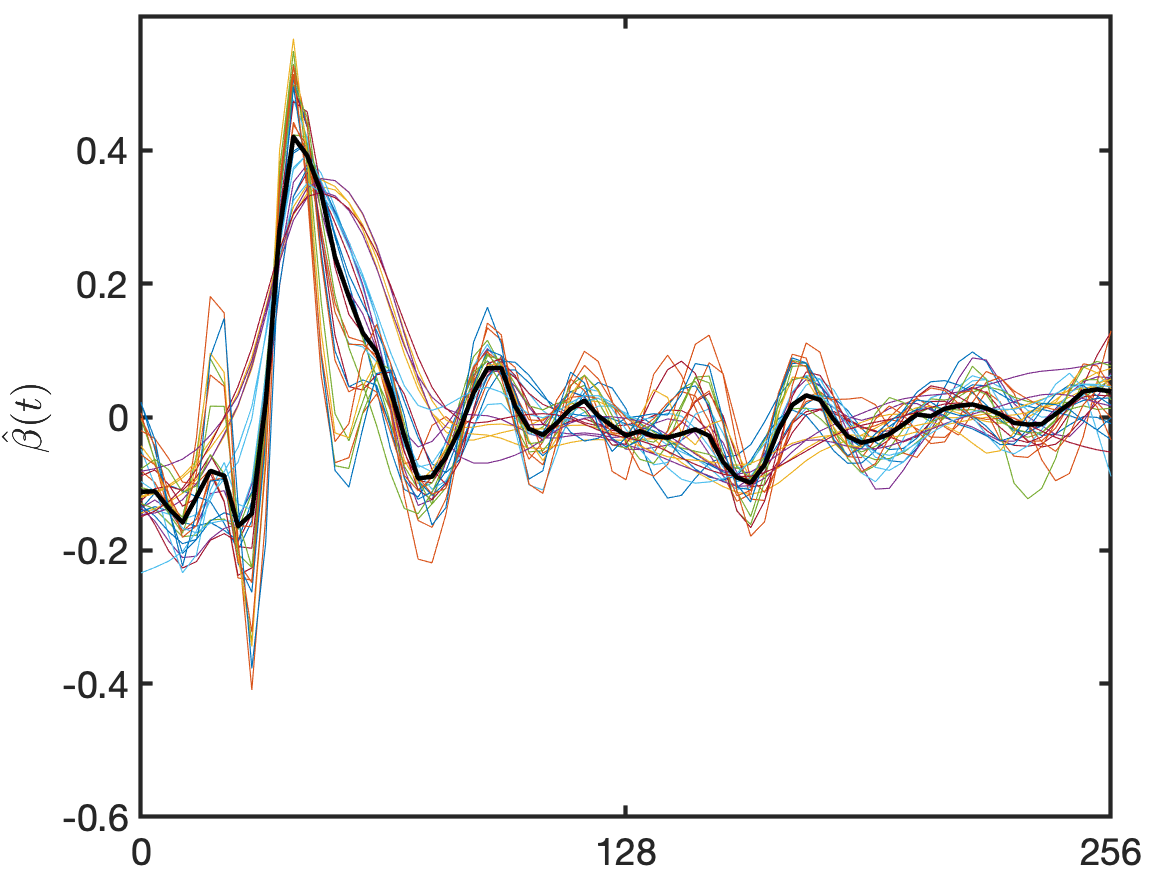}
        \caption*{FLDA}
        \end{subfigure}       
        \begin{subfigure}[b]{0.245\textwidth}
        \centering
        \includegraphics[width = \textwidth, height = 1in]{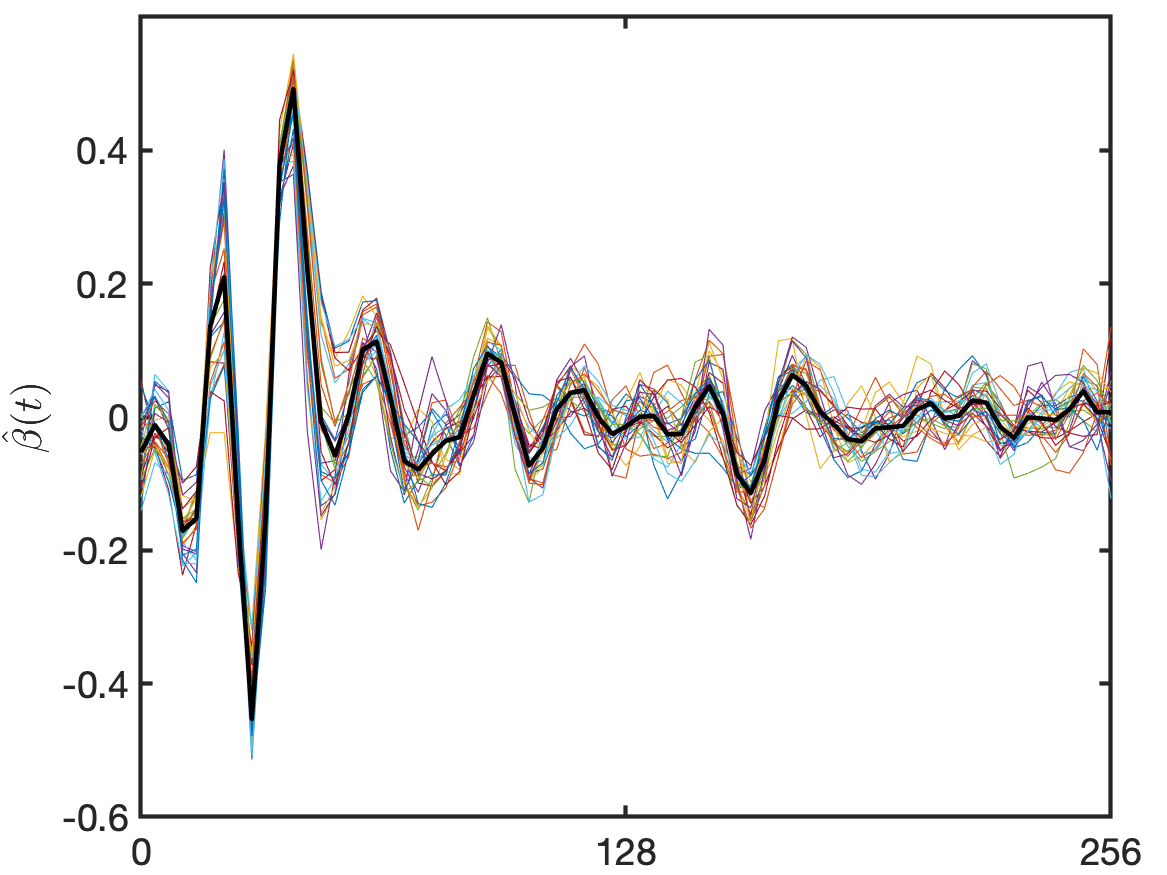}
        \caption*{RFLDA}
        \end{subfigure}       
        \begin{subfigure}[b]{0.245\textwidth}
        \centering
        \includegraphics[width = \textwidth, height = 1in]{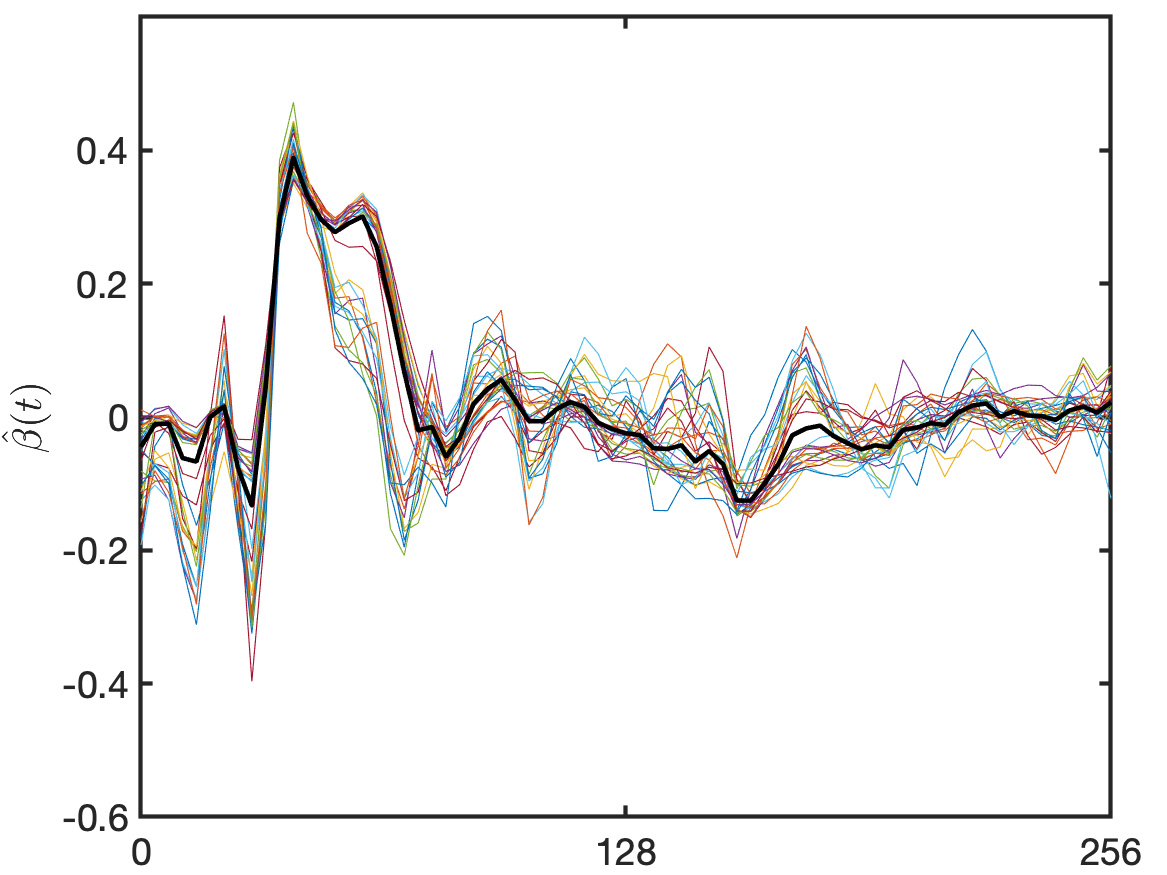}
        \caption*{PLS}
        \end{subfigure}       			
        }   
\end{figure}

\begin{table}[bp]
\begin{center} 
\caption{\label{tab:realerr} Mean test errors (\%) from real data examples, based on 30 repetitions. Standard errors are in the parentheses.}
\footnotesize{
\begin{tabular}{cccccccccc}
Data &  SFLDA & FLDA & RFLDA & PLS & B-Gauss & B-NPD & B-NPR & Logistic & QDA \\ 
\multirow{ 2}{*}{Tecator} &   5.0  &  3.4  &  7.4  &  4.8  &  6.3  &  4.2  &  4.8  &  1.8 &   3.3 \\
&  (0.44)  &  (0.41)  &  (0.68)  &  (0.46)  &  (0.48)  &  (0.42) &   (0.59)  &  (0.27)   & (0.42)\\
\multirow{ 2}{*}{Wine} &   10.3  &  10.0  &  9.2  &  9.2 &   11.1  &  10.4  &  9.9 &   8.7  &  11.6 \\
&  (0.81) &   (0.74)  &  (0.72)  &  (0.81) &   (0.91)  &  (0.89)  &  (0.57)  &  (0.65)  &  (0.80) \\
\multirow{ 2}{*}{Growth} &  6.0  &  5.7  &  8.3  &  6.1   & 6.2  &  7.2  &  7.1  &  5.9  &  6.7 \\
& (0.39)  &  (0.39)  &  (0.43)  &  (0.32)  &  (0.44)  &  (0.36)  &  (0.40) &   (0.48)  &  (0.40)\\
\multirow{ 2}{*}{dGrowth} &   7.0  &  6.4  &  7.4  &  6.3  &  7.4  &  8.1 &   7.6  &  7.1 &   9.5 \\
&(0.57)   & (0.47)   & (0.52)  &  (0.52)  &  (0.48)   & (0.49)  &  (0.55) &   (0.56)  &  (0.60) \\
\multirow{ 2}{*}{Wheat} & 0.1  &  0.0  &  0.1   &      0.0   & 0.1   &      0.0  &  0.2  &  0.2  &  0.2 \\
&(0.09)    &(0.00)    &(0.33)         &(0.00)    &(0.09)         &(0.00)    &(0.13)    &(0.17)    &(0.13) \\
\multirow{ 2}{*}{Phoneme} & 19.7 & 21.6 & 20.9 & 21.4 & 23.7 & 23.8 & 24.2 & 22.1 & 24.4 \\
&(0.53) & (0.59)   &  (0.53)   & (0.58) &    (0.60)  & (0.57)  &  (0.60)   & (0.62) &    (0.65) \\
\end{tabular}}
\end{center}
\end{table}

We tested the same set of the functional classification methods as in the previous section for real data examples. The Tecator dataset \citep{DelaigleHall2012} has 240 curves of near infrared absorbance spectra (850 - 1050 nm) of finely chopped meat, using a Tecator Infratec Food \& Feed Analyser. Here the groups are defined according to the fat content. The Wine spectra dataset \citep{DaiMuellerYao2017} contains 123 samples of mid infrared spectra (4000 - 400 cm$^{-1}$), which are divided into two groups based on the alcohol content level. The Growth dataset \citep{GasserPrader1984, Sheehy1999} has height growth curves of 112 boys and 120 girls from births to the 18th year. We also analysed the velocity of the Growth curves. The Wheat data \citep{kalivas1997two} have near infrared spectra of 100 wheat samples, divided into two groups according to the protein content. The Phoneme data are log-periodograms constructed from digitized speech of two different sounds ``aa" and ``ao", as described in \citet{Hastie2009}.
Each panel of Figure \ref{fig:realdata} respectively shows 40 randomly selected curves from each data set, with each class shown with different colors. 

To estimate test errors, we split into 2/3 training and 1/3 testing data, and tune each method in the same way as the simulation study, and repeat this process 30 times. Mean test errors are displayed in Table \ref{tab:realerr}. While there are no meaningful differences among the test errors of these methods, we can see that the proposed method achieves compatible accuracies with a much better ability to find localized discriminant regions, as can be seen in Figure \ref{fig:realbeta}. In particular, we have found that the selected regions of Tecator, Wine, and Phoneme are consistent with findings from the spectroscopy literature. Specifically, according to the body composition study \citep{conway1984new}, fat has high absorbance around 930nm, the absorbance region for ethanol is concentrated in 1200 - 850 cm$^{-1}$ \citep{debebe2017non}, and it has been found by \citet{hastie1995penalized} that the discriminating feature in phonemes are in the low frequencies about 500 - 1000Hz, corresponding to the frequencies 16--32. All these known regions are consistent with the estimates from the proposed method in the figure.

\section*{Acknowledgment}

We are grateful to Aurore Delaigle and Xiongtao Dai for sharing the Matlab codes for compared methods. We are also grateful to two referees, an associate editor and the editor for careful reading and helpful suggestions. 
The third author's work was supported by Basic Science Research Program of the National Research Foundation of Korea (NRF-2019R1A2C1005979) funded by the Korean government.

\section*{Supplementary material}
Supplementary material 
includes additional details on relevant background material and proofs, as well as an example of matlab codes for implementation.

\bibliography{funLDA_ref}

\end{document}